\documentclass[fleqn,usenatbib]{mnras}

\usepackage[T1]{fontenc}

\DeclareRobustCommand{\VAN}[3]{#2}
\let\VANthebibliography\thebibliography
\def\thebibliography{\DeclareRobustCommand{\VAN}[3]{##3}\VANthebibliography}

\usepackage{amsmath}
\usepackage[varg]{txfonts}
\pdfoutput=1
\usepackage{graphicx}
\usepackage{newtxtext,newtxmath}

\usepackage{textcomp}
\usepackage[inline]{enumitem}
\usepackage{hyperref}
\usepackage{float}
\usepackage{placeins}
\usepackage{natbib}

\usepackage{amsmath, amssymb, amsfonts}
\usepackage{cmdDef}
\usepackage{xcolor}
\usepackage[normalem]{ulem} 
\usepackage{booktabs}
\usepackage{CJKutf8}

\newcommand\hl{\bgroup\markoverwith
  {\textcolour{yellow}{\rule[-.5ex]{2pt}{2.5ex}}}\ULon}

\title[Assembly of NGC~3957]{The GECKOS survey: Assembly history of the lenticular galaxy NGC~3957}

\author[Y. Ding et al.]{\begin{CJK*}{UTF8}{gbsn}Y. Ding (丁郁琛)\end{CJK*}$^{1}$\thanks{E-mail: y.ding@ljmu.ac.uk}, 
M. Martig$^{1}$,
L. Zhu$^{2}$,
A. Fraser-McKelvie$^{3}$,
R. Leaman$^{4}$,
G. van de Ven$^{4}$,
\newauthor
J. van de Sande$^{5}$,
F. Pinna$^{6,7}$,
E. Emsellem$^{3,8}$,
F. Fragkoudi$^{9}$,
Y. Jin$^{10}$,
A. Poci$^{11}$,
C. de S\'a-Freitas$^{12}$,
\newauthor
M. Frosst$^{13}$,
R. Cai$^{2,14}$,
S. M. Croom$^{15}$,
T. A. Davis$^{16}$,
R. Elliott$^{17}$,
M. R. Hayden$^{18}$,
J. Falc\'on-Barroso$^{6,7}$,
\newauthor
D. A. Gadotti$^{19}$,
A. Marasco$^{20}$,
L. M. Valenzuela$^{21}$,
Z. Wang (Purmortal)$^{22}$,
E. Wisnioski$^{23}$,
M. Yang$^{2}$,
\newauthor
T. Zafar$^{24,25}$
\\
$^{1}$Astrophysics Research Institute, Liverpool John Moores University, 146 Brownlow Hill, Liverpool L3 5RF, UK\\
$^{2}$Shanghai Astronomical Observatory, Chinese Academy of Sciences, 80 Nandan Road, Shanghai 200030, China\\
$^{3}$European Southern Observatory, Karl-Schwarzschild-Stra{\ss}e 2, Garching, 85748, Germany\\
$^{4}$Department of Astrophysics, University of Vienna, T{\"u}rkenschanzstra{\ss}e 17, 1180 Vienna, Austria \\
$^{5}$School of Physics, University of New South Wales, Sydney, NSW 2052, Australia\\
$^{6}$Instituto de Astrof\'isica de Canarias, Calle V\'ia L\'actea s/n, E-38205. La Laguna, Tenerife, Spain\\
$^{7}$Departamento de Astrof\'isica. Universidad de La Laguna, Av. del Astrof\'isico Francisco S\'anchez s/n, E-38206, La Laguna, Tenerife, Spain\\
$^{8}$Univ Lyon, Univ Lyon1, ENS de Lyon, CNRS, Centre de Recherche Astrophysique de Lyon UMR5574, F-69230 Saint-Genis-Laval France \\
$^{9}$Institute for Computational Cosmology, Department of Physics, University of Durham, South Road, Durham DH1 3LE, UK\\
$^{10}$Department of Astronomy, Westlake University, Hangzhou, Zhejiang 310030, China\\
$^{11}$Sub-department of Astrophysics, Department of Physics, University of Oxford, Denys Wilkinson Building, Keble Road, Oxford OX13RH, UK \\
$^{12}$European Southern Observatory, Alonso de Córdova 3107, Vitacura, Región Metropolitana, Chile\\
$^{13}$International Centre for Radio Astronomy Research (ICRAR), University of Western Australia, Crawley, WA 6009, Australia\\
$^{14}$School of Astronomy and Space Sciences, University of Chinese Academy of Sciences, No. 19A Yuquan Road, Beijing 100049, China\\
$^{15}$Sydney Institute for Astronomy, School of Physics, University of Sydney, NSW 2006, Australia\\
$^{16}$Cardiff Hub for Astrophysics Research \&\ Technology, School of Physics \&\ Astronomy, Cardiff University, Queens Buildings, Cardiff, CF24 3AA, UK\\
$^{17}$Centre for Astrophysics and Supercomputing, Swinburne University of Technology, Hawthorn, VIC 3122, Australia\\
$^{18}$Homer L. Dodge Department of Physics \& Astronomy, University of Oklahoma, 440 W. Brooks St., Norman, OK 73019, USA \\
$^{19}$Centre for Extragalactic Astronomy, Department of Physics, Durham University, South Road, Durham DH1 3LE, UK\\
$^{20}$INAF – Osservatorio Astronomico di Padova, Vicolo dell’Osservatorio 5, I-35122 Padova, Italy\\
$^{21}$Universit\"ats-Sternwarte, Fakult\"at f\"ur Physik, Ludwig-Maximilians-Universit\"at M\"unchen, Scheinerstr. 1, 81679 M\"unchen, Germany\\
$^{22}$Department of Physics and Astronomy, University of Utah, Salt Lake City, UT 84112, USA \\
$^{23}$Research School of Astronomy and Astrophysics, Australian National University, Canberra, Australia \\
$^{24}$School of Mathematical and Physical Sciences, Macquarie University, NSW 2109, Australia\\
$^{25}$Astrophysics and Space Technologies Research Centre, Macquarie University, Sydney, NSW 2109, Australia\\
}

\date{Accepted XXX. Received YYY; in original form ZZZ}

\pubyear{2026}

\begin{document}
\label{firstpage}
\pagerange{\pageref{firstpage}--\pageref{lastpage}}
\maketitle

\begin{abstract}
We analyse the assembly history of the edge-on lenticular galaxy NGC~3957 using deep integral-field spectroscopic MUSE data from the GECKOS survey. By applying a dust-corrected Multi-Gaussian Expansion and a population-orbit superposition model, we disentangle the galaxy's stellar kinematics, age, and metallicity. We dynamically decompose the galaxy and identify three distinct components: a dynamically-cold main disc, a compact Nuclear Stellar Disc (NSD), and a hot component. The NSD emerges as the youngest and most metal-rich component ($t = 6.9 \pm 0.4$~Gyr; $[Z/H] = 0.49 \pm 0.06$~dex), implying that the stellar bar is a long-lived structure that formed at least $\sim 7$~Gyr ago. The main stellar disc is dynamically cold ($\sigma_z \sim 20-30$~km/s), precluding any significant mergers over the last $\sim 8$~Gyr, and exhibits a strong positive age gradient (younger inside, older outside) beyond the bar radius. Synthesising these dynamical fossil records, NGC~3957 likely evolved as a `faded spiral' in a small-to-medium group environment. Its outer disc might passively fade due to mild gas starvation, while the bar fuelled prolonged central star formation. Comparison with S0s in the Fornax cluster reveals that this combination of internal secular evolution and mild starvation produces `outside-in' fading signatures that could mimic the environmental stripping typically seen in dense clusters.
\end{abstract}




\section{Introduction}\label{sec:intro}
Lenticular galaxies (S0s) are a ubiquitous galaxy population in the local Universe, yet their exact formation pathways remain a subject of active debate. The transformation from star-forming spiral galaxies to quiescent S0s is generally attributed to a combination of environment-driven processes (`nurture') and internal secular evolution (`nature'). In dense environments such as massive galaxy clusters, external mechanisms like ram-pressure stripping \citep{gunn1972} are highly efficient at rapidly removing gas and quenching star formation \citep[e.g.,][]{dressler1980,bekki2002}. In less dense environments, such as loose groups or the field, the supply of fresh cosmological gas may simply be cut off \citep[strangulation, e.g.,][]{balogh2000,Dekel2006,kawata2008,Peng2015}. As the remaining gas is consumed, the young, bright stellar populations die out, causing the extended disc to passively fade. This passive evolution naturally increases the apparent bulge-to-disc luminosity ratio without dynamically growing the bulge, producing a quiescent S0 galaxy commonly referred to as a `faded spiral' \citep[e.g.,][]{Falcon-Barroso2015,Croom2021}.

In such field or group environments, galaxy mergers also represent a prominent formation channel for S0s \citep[e.g.,][]{Deeley2021, Coccato2022}. Mergers can drive morphological transformation by kinematically heating the stellar disc, growing the central bulge, and triggering central starbursts that rapidly deplete gas reservoirs \citep[e.g.,][]{Bekki1998, Bournaud2005}. While it is often assumed that mergers destroy rotating discs \citep[e.g.][]{Villalobos2008,Pablo2023}, simulations have shown that merger remnants can still retain significant rotation. In particular, gas-rich mergers can allow a dynamically cold, thin disc to rapidly reform via the cooling of the accreted gas \citep{Yu2023}, successfully reproducing the kinematic properties of fast-rotating S0s observed in large integral-field spectroscopic (IFS) surveys \citep[e.g.,][]{querejeta2015}.

In the absence of frequent high-speed encounters, internal secular processes shape the evolution of field and group galaxies. Stellar bars, for instance, play a crucial role in redistributing angular momentum and driving gas inflows towards the galactic centre \citep[e.g.,][]{athanassoula2003, Kormendy2004}. This continuous funnelling of gas can fuel prolonged star formation, potentially building compact central structures like Nuclear Stellar Discs \citep[NSDs;][]{Schultheis2025} while the outer disc may gradually fade. Consequently, the evolutionary history of field S0s remains complex, heavily dependent on their specific assembly histories rather than their environment alone \citep{Deeley2021, Croom2021}. For instance, spatially-resolved studies suggest that the distinct structural components within S0s follow different formation pathways: dynamically hotter and thicker components are predominantly built via early mergers, whereas dynamically cold thin discs are largely shaped by prolonged secular evolution \citep[e.g.,][]{Yang2021}. To fully reconstruct their evolutionary pathways, it is helpful to dynamically disentangle their internal orbital components—specifically differentiating between bulges, secularly built NSDs, and bars.

Disentangling these components observationally, however, poses a significant challenge. Detailed studies characterizing the star formation histories of central substructures often favour face-on or low-inclination systems, where bars and NSDs are morphologically distinct \citep[e.g., the TIMER and PHANGS surveys;][]{Gadotti2019, Bittner2020, Pessa2023}. On the other hand, resolving the detailed vertical structure and dynamical heating history of galactic discs specifically requires an edge-on perspective \citep[e.g.,][]{Comeron2011,pinna2019,pinna2019a,martig2025}. Pioneering massive IFS surveys like ATLAS$^{\rm 3D}$ \citep{cappellari2011}, CALIFA \citep{sanchez2012}, SAMI \citep[e.g.,][]{vandeSande2017}, and MaNGA \citep[e.g.,][]{Graham2018} have revolutionised our understanding of the overall kinematic dichotomy of early-type galaxies, physically classifying them into fast and slow rotators based on their stellar angular momentum \citep[e.g.,][]{emsellem2007,Emsellem2011,cappellari2016}. However, the majority of these large-scale surveys are limited to kiloparsec-scale spatial resolutions, at which compact central structures like NSDs (typically within a few 100 pc) cannot be resolved. Furthermore, studying these internal structures in edge-on systems remains exceptionally difficult due to dust obscuration and the severe blending of overlapping stellar structures along the line of sight \citep[e.g.,][]{Rutherford2025}. This highlights the critical need for deep, high-resolution observations combined with sophisticated dynamical decomposition.

To overcome the degeneracy caused by this line-of-sight integration, advanced dynamical modelling is essential. While various techniques exist, such as Jeans-Anisotropic-MGE \citep[JAM; e.g.,][]{Cappellari2008,Li2017} and Made-to-Measure models \citep[M2M; e.g.,][]{Syer1996,deLorenzi2007,Long2010,Long2012,Zhu2014}, the Schwarzschild orbit-superposition method \citep{schwarzschild1979} is particularly powerful. By reconstructing the galaxy's orbital backbone without ad-hoc assumptions about the distribution function, the triaxial implementation of this method \citep[e.g.,][and the publicly available \texttt{DYNAMITE} code; \citealt{jethwa2020,Thater2022}]{vandenbosch2008} has been extensively validated in simulations \citep[e.g.,][]{zhu2018a, jin2019} and applied to uncover the intrinsic 3D structures of galaxies across various geometries \citep[e.g.,][]{Zhu2018c,jin2020}. 

Crucially, this modelling technique has undergone significant enhancements in recent years. It has been updated to explicitly include non-axisymmetric bar structures \citep[e.g.,][]{Tahmasebzadeh2022,Tahmasebzadeh2024,jin2025a}, and simultaneously evolved into a population-orbit superposition method \citep[e.g.,][]{poci2019, zhu2020, Ding2023, jin2024}. By tagging orbits with specific ages and metallicities, the model can simultaneously fit stellar kinematic and population maps from IFS data. This state-of-the-art framework enables a physically motivated chemo-dynamical decomposition, allowing us to successfully isolate overlapping components (e.g., discs, bars, and bulges) and reconstruct their distinct assembly histories \citep[e.g.,][]{Zhu2022a, vdv2025, jin2025b}.

In this work, we use the population-orbit superposition method to perform a chemo-dynamical decomposition of NGC~3957, a typical edge-on S0 galaxy. Throughout this paper, we adopt a flat $\Lambda$CDM cosmology with $\Omega_m = 0.30$, $\Omega_\Lambda = 0.70$, and $H_0 = 70 \ \mathrm{km\ s^{-1}\ Mpc^{-1}}$, and a Chabrier \citep{Chabrier2003} initial mass function (IMF).

The paper is organized as follows. We introduce the dataset in Section \ref{sec:data}. We describe the population-orbit superposition model and its relevance to the orbital decomposition in Section \ref{sec:method}. We show the results of orbital decomposition and the assembly history in Section \ref{sec:result}. We discuss our results in Section \ref{sec:discussion} and present our summary in Section \ref{sec:conclusion}.

\section{Sample and Data}\label{sec:data}
\subsection{The GECKOS Survey}
NGC~3957 is analysed as part of the second internal data release (iDR2) of the GECKOS survey, which aims at investigating the evolution of disc galaxies by observing 36 Milky Way-mass edge-on galaxies ($i>85^\circ$) in the local Universe ($10<D<70$~Mpc). 
The survey uses the MUSE instrument in wide-field mode to provide integral field spectroscopy with high spatial resolution ($0.2^{\prime\prime}$ pixel scale) and a deep surface brightness limit extending to $\mu_V = 23.5$~mag~arcsec$^{-2}$. By stacking, in some cases, several hours' worth of observations of faint disc regions, the survey reaches these impressive detection limits.
Unlike previous surveys, GECKOS is specifically designed to resolve the vertical structure of discs with high physical resolution of $<200$~pc \citep[providing diagnostics of kinematic sub-structures, see for example,][] {Fraser-McKelvie2025,Rutherford2025} and deep sensitivity to low surface brightness features. This enables the detection of faint extraplanar features and outflows \citep{ciraulo2025,elliott2026}. 

The raw data are reduced and mosaicked using the \texttt{pymusepipe} package \citep{Emsellem2022}, which wraps the standard ESO MUSE pipeline \citep{Weilbacher2020} and \texttt{esorex} recipes. The detailed setup of the data reduction pipeline is described in \citet{Fraser-McKelvie2025} and more details will be presented in van de Sande et al. (in prep.).

\subsection{Target Galaxy: NGC~3957}
NGC~3957, located at a distance of 24.8~Mpc, serves as a prime example for studying the vertical structure and assembly history of S0 galaxies \citep{buta2015} within the GECKOS sample. It is a satellite galaxy residing in the small-to-medium NGC~4038 galaxy group \citep[LGG 263; which consists of 8-26 member galaxies,][]{garcia1993}. It has a stellar mass of $10^{10.55}\,\mathrm{M}_\odot$ (van de Sande et al., in prep). Morphologically, it is characterised by a prominent boxy/peanut-shaped (B/P) bulge clearly visible in near-infrared imaging, a well-known signature of a bar viewed edge-on \citep{lutticke2000}. Previous photometric decompositions have also revealed a thin and a thick disc co-existing in this galaxy \citep{pohlen2004}.

Visually, NGC~3957 features a prominent equatorial dust lane, indicating the presence of an interstellar medium. The existence of such cold gas and dust in early-type galaxies is not unusual; for instance, the ATLAS$^{\rm 3D}$ survey revealed that cold molecular gas is present in $\sim 22\%$ of local early-type galaxies \citep{Young2011}. However, the galaxy has a low star formation rate ($\mathrm{SFR} \approx 0.26\,\mathrm{M}_\odot \mathrm{yr}^{-1}$, van de Sande et al., in prep). In fact, compared to the broader GECKOS sample, NGC~3957 possesses one of the lowest SFRs.
This relatively lower level of dust obscuration and active star formation allows us to extract much cleaner stellar kinematics compared to more gas-rich spirals, making it an ideal target for constructing a robust dynamical model. Consistent with this relative quiescent nature, the ionised gas (H$\alpha$) emission is extremely faint ($10^{-16} \sim 10^{-17}\ \rm{erg\ s^{-1}\ cm^{-2}}$) and mostly concentrated within 2 kpc from the disc \citep{elliott2026}. The main properties of NGC~3957 are listed in Table~\ref{tab:galaxy_prop}.

\begin{figure}
    \centerline{
        \includegraphics[width=1.\columnwidth]{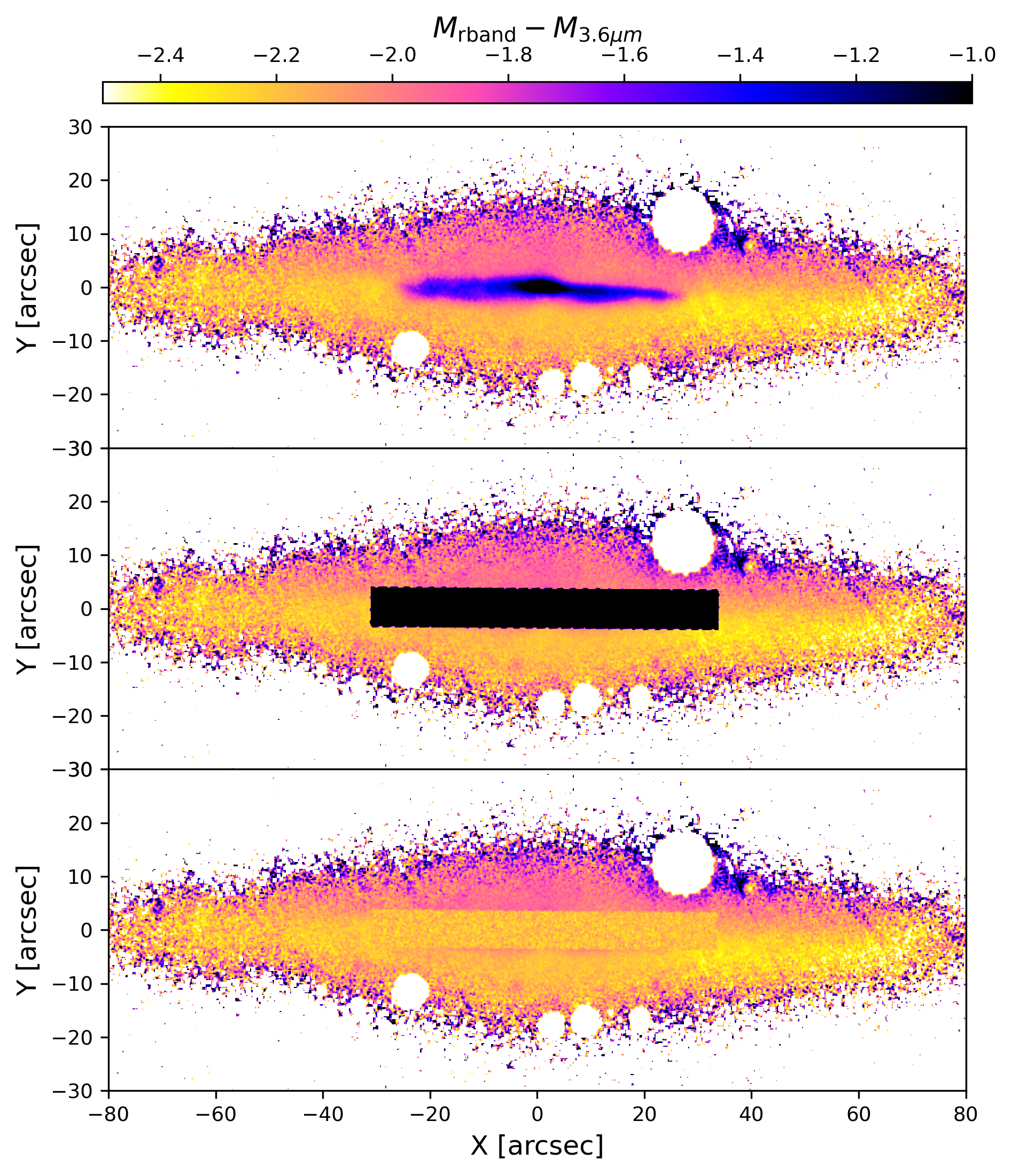}}
    \caption{
    Illustration of the dust correction strategy for the MUSE $r$-band image. {\em (Top panel:)\/} The $M_\mathrm{rband}-M_{3.6 \mu m}$ colour map, where the dark blue band along the major axis indicates regions of significant dust attenuation. We identify the dust-affected area using a colour cut of  $M_\mathrm{rband}-M_{3.6 \mu m}>-1.2$ 
    {\em (Middle panel:)\/} The rectangular mask defined to cover the prominent dust lane. {\em (Bottom panel:)\/} The colour-corrected $r$-band image. The flux within the masked region has been substituted using the scaled flux from the $3.6 \mu m$ image combined with the average colour $\bar{C}$ measured outside the mask, ensuring a smooth surface brightness profile for MGE fitting.
    }
    \label{fig:color_map}
\end{figure}

\begin{figure}
    \centerline{
        \includegraphics[width=1.\columnwidth]{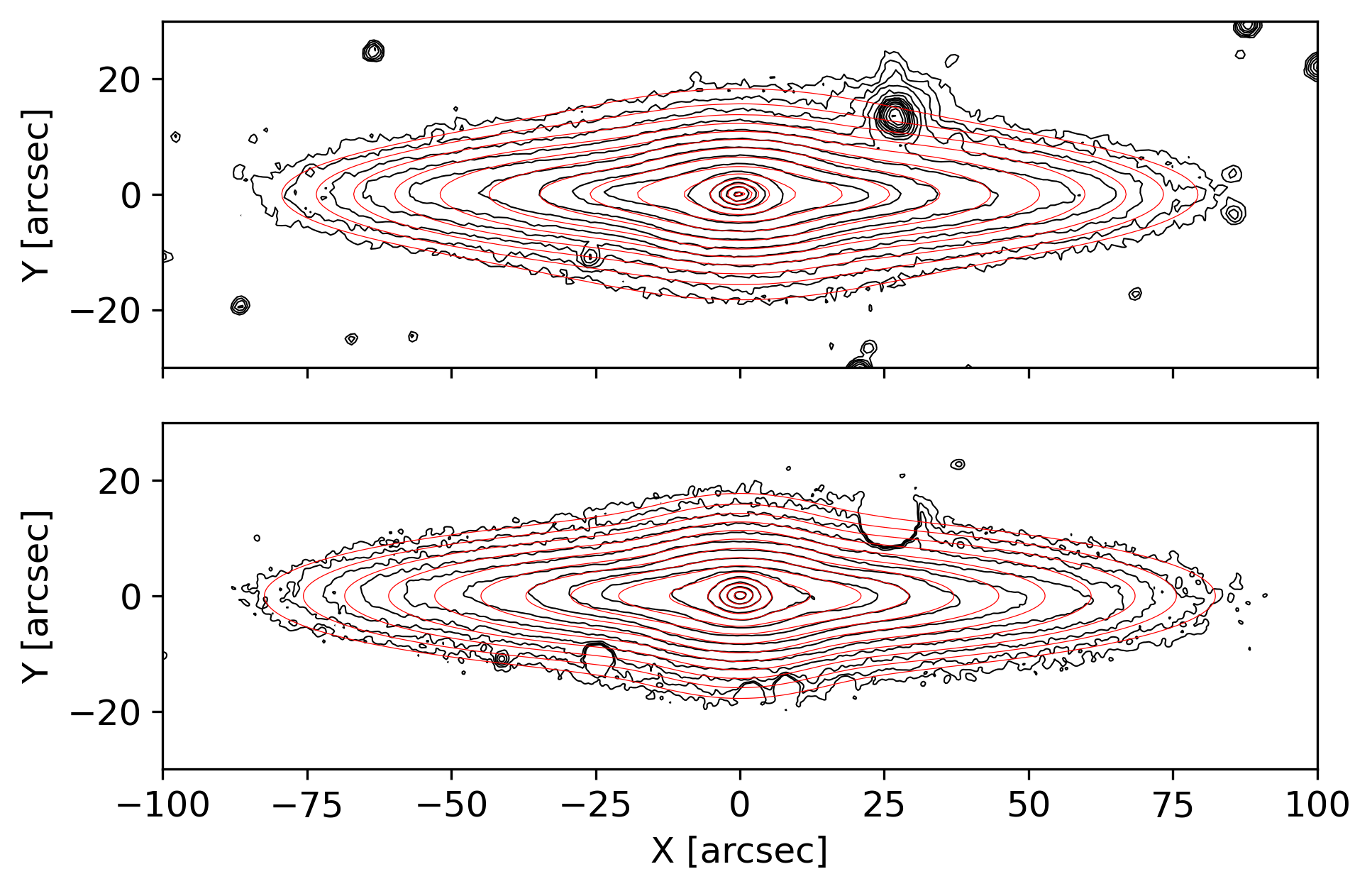}}
    \caption{MGE fit to NGC~3957. {\em (top panel:)\/} Spitzer 3.6µm imaging contours are in black, the MGE model contours are in red. {\em (bottom panel:)\/} the MGE model contours based on the dust-corrected $r$-band image.
    }
    \label{fig:MGE_contour_mix}
\end{figure}

\begin{table}
   \centering
   \small 
   \setlength{\tabcolsep}{3.5pt} 
   \renewcommand\arraystretch{1.5}
   \centering
   \begin{tabular}{cccccc}
   \hline\hline
   Distance & $R_{\mathrm{e}}$ & $M_\star$ & SFR & PA & $i$ \\
        {[Mpc]} &[arcsec] & [$\log_{10} M_\star/{\rm M}_{\odot}$] & [${\rm M}_\odot/{\rm yr}$] & [$^{\circ}$] & [$^{\circ}$] \\
     (1)  &  (2) & (3) & (4) & (5) & (6) \\
   \hline
    24.8 & 22.2 & 10.55 & 0.26 & 173 & 88$\pm$1\\ 
   \hline
   \end{tabular}
   \caption{Properties of NGC~3957.  The distance (1) is from \citet{tully2023}, the effective radius (2),  is calculated from MGE of the dust-corrected $r$-band image. Stellar mass (3) and SFR (4) are from van de Sande et al (in prep). The position angle (5) is determined by photometric isophotes and adopted for the MGE fitting. The inclination angle (6) is from \citet{pohlen2004}.
   }\label{tab:galaxy_prop}
   \end{table}
   
\subsection{Stellar kinematics and stellar population maps}
We derive the stellar kinematics and stellar population properties using the \texttt{nGIST} pipeline \citep{Fraser-McKelvie2025}\footnote{\url{https://ascl.net/2507.015}}. The \texttt{nGIST} pipeline prepares and organises galaxy IFS data cubes, spatially bins spectra to a specified signal-to-noise ratio (S/N), and wraps commonly-used spectral analysis tools to analyse the data in a flexible and modular manner. To ensure the reliable extraction of higher-order kinematic moments, the data cubes are spatially binned using the Voronoi tessellation method \citep{cappellari2003} to achieve a target S/N of 100. This S/N is calculated over the restricted wavelength range of 4800--7000~\AA\ to minimize the impact of strong sky lines in the redder part of the spectrum. For NGC~3957, the spatial resolution is well-preserved with typical bin sizes of $\sim 20$~pc in the central regions, while in the outskirts, the bins dynamically expand up to $\sim 600$~pc to meet the target S/N.

The stellar line-of-sight velocity distribution (LOSVD) and stellar population parameters (age and metallicity) are extracted using \texttt{nGIST}'s KIN and SFH modules, respectively.  
Both modules use the Penalized Pixel-Fitting (\texttt{pPXF}) code \citep{cappellari2004, Cappellari2017b}. For the kinematic extraction, \texttt{pPXF} fits the spectra using a $23^{\text{rd}}$-order additive Legendre polynomial to match the templates to the data, alongside a $1^{\text{st}}$-order multiplicative polynomial to account for minor continuum variations caused by imperfect sky subtraction and dust attenuation. To derive the stellar population maps, \texttt{pPXF} determines the star formation history by fitting the observed spectra as a non-negative linear combination of Simple Stellar Population (SSP) templates, while simultaneously applying a \citet{Calzetti2000} extinction curve. This effectively rectifies the continuum shape and ensures that the extracted stellar properties are not severely biased by dust reddening.

The kinematic spectral fitting was performed over the broad wavelength range of 4800--8950~\AA, using the \texttt{xsl\_stars} template library \citep[X-shooter Spectral Library;][see Section 2.4 in \citealt{Fraser-McKelvie2025}]{Verro2022}. This extended spectral range includes key absorption features such as the Calcium II triplet (CaT) near $\sim 8500$~\AA. Because redder wavelengths are significantly less affected by dust extinction than optical bands, incorporating the CaT region is crucial for robustly penetrating the dust lane and accurately constraining the velocity dispersion in dynamically cold, dusty edge-on discs. The resulting 2D maps of mean velocity $V$, velocity dispersion $\sigma$, and Gauss-Hermite moments $h_3$ and $h_4$ \citep{Gerhard1993,Rix1997} serve as the primary inputs for our dynamical models.

\subsection{Multi-Gaussian expansion}\label{sec:MGE}
In order to build up the dynamical model of NGC~3957, we use the Multi-Gaussian Expansion \citep[MGE;][]{cappellari2002} to fit the galaxy image.
To constrain the 3D gravitational potential, we require an unbiased estimate of the projected mass distribution, which is best given by the near-infrared $3.6\mu$m Spitzer IRAC image \citep{Sheth2010} as it is minimally affected by dust \citep[e.g.,][]{mcgaugh2014}. Conversely, the 2D luminosity density is reproduced by weighting the orbits to fit the observed MUSE kinematics. Since these kinematics are extracted from optical spectra, the corresponding surface brightness constraints must also be defined in the optical regime to ensure physical consistency. Therefore, the $r$-band photometry is the most appropriate choice to represent this kinematic tracer population. However, unlike the near-infrared data, the optical $r$-band image is heavily obscured by a prominent dust lane along the major axis. Thus, we need a careful dust solution \citep[see also][for recent dust-corrected dynamical modelling]{Rutherford2025} before performing the MGE fit.

In both the $3.6\mu$m and $r$-band MGE fits, we consider an axisymmetric solution by adopting a constant position angle (PA) for the Gaussians to match the photometric signature of the inner or outer disc.
To perform the spatially-resolved dust correction, we use the Pan-STARRS $r$-band image \citep{Chambers2016} and the $3.6\mu$m image. We specifically opt for the Pan-STARRS r-band image rather than the one extracted from the MUSE cube because its larger spatial coverage perfectly matches that of the 3.6$\mu$m data, which is essential for constructing a robust, global MGE model. To properly combine these two datasets, both images must be accurately aligned on a common pixel grid. The native pixel scales are distinctly different: $0.25^{\prime\prime}$/pixel for the Pan-STARRS image and $0.75^{\prime\prime}$/pixel for the $3.6\mu$m image. To address this, we use the World Coordinate System (WCS) from the image headers to map the exact celestial footprint of the $r$-band image onto the $3.6\mu$m data. We first crop the $3.6\mu$m image to match the identical field of view. Subsequently, we resample the lower-resolution $3.6\mu$m image to match the fine $0.25^{\prime\prime}$/pixel grid (resulting in matching $1440 \times 1440$ pixel arrays) using a bilinear interpolation algorithm. During this resampling process, the total flux of the $3.6\mu$m image is strictly conserved. Once spatially aligned and resolution-matched, both images are converted into the magnitude space to construct the $M_\mathrm{rband}-M_{3.6 \mu m}$ colour map for identifying and masking the dust lanes.

The correction procedure is as follows: (a) we first evaluate the region where the dust influence is most significant by plotting the colour map, i.e., the magnitude difference $C = M_\mathrm{rband}-M_{3.6 \mu m}$  (see the blue region in the top panel of Fig.~\ref{fig:color_map}). 
(b) We then define a rectangular mask covering the central dust lane. We choose a simple rectangular mask rather than a strict contour-based mask (e.g., exactly outlining the $-1.2$ Mag threshold) because the actual dust distribution is highly irregular. A contour-based mask creates highly non-elliptical boundaries, and the MGE fitting is notoriously sensitive to such irregularities. These jagged edges would dominate the local $\chi^2$ and severely degrade the global MGE fit. A conservative rectangular mask avoids this issue and provides a much more stable boundary for the fit (Fig.~\ref{fig:color_map}, middle panel). 
(c) We calculate the average colour $\bar{C} = \langle M_\mathrm{rband}-M_{3.6 \mu m} \rangle$ and its standard deviation $\sigma_C$ using pixels outside this masked region. 
(d) We then replace the pixel values within the masked region of the $r$-band image. The corrected magnitude at each pixel inside the mask is estimated as $M_{\mathrm{rband,corr}} = M_{3.6\mu \mathrm{m}} + \bar{C}$. 
(e) To account for potential variations in the intrinsic colour and the substitution process, we apply a Gaussian uncertainty to the replaced flux based on the measured colour dispersion $\sigma_C$.

Finally, we perform the MGE fitting separately on the $3.6\mu$m image and the dust-corrected $r$-band image (Fig.~\ref{fig:color_map}, bottom panel). Fig.~\ref{fig:MGE_contour_mix} shows the resulting MGE contours overlaid on both the $3.6\mu$m (top panel) and the $r$-band images (bottom panel). A detailed comparison of the MGE fit to the dust-corrected $r$-band data and its residuals is demonstrated in Fig.~\ref{fig:MGE_fitting} in the Appendix. The best-fitting MGE parameters for both images are listed in Tables~\ref{tab:MGE_Mass} and \ref{tab:MGE_Light} in the Appendix.

\section{Population-orbital superposition method}\label{sec:method}
In this study, we apply the population-orbit superposition method \citep{zhu2020} to reconstruct the intrinsic properties of NGC~3957. This technique allows us to fit the luminosity density, kinematic, age, and metallicity maps. Once the best-fitting model is determined, we extract the internal stellar orbit distribution associated with the age and metallicity distributions.
Based on the derived orbital structure in the phase space of circularity and radius, we decompose the galaxy into distinct dynamical components. In this paper, we identify three dynamical components: a dynamically-cold main disc, a central NSD, and a dynamically-hot component. Subsequently, we extract the face-on surface brightness, age, and metallicity radial profiles for each component separately. We illustrate the whole process in Fig.~\ref{fig:model_fitting} and Fig.~\ref{fig:phase_space_marked} for NGC~3957.

Briefly, the modelling procedure involves four main stages: (i) constructing the model of the gravitational potential, (ii) calculating the orbital library under the gravitational potential, (iii) determining the orbital weights fitting the luminosity density and kinematic maps, and (iv) tagging the weighted orbits with age and metallicity. Step (i)-(iii) are performed with the \texttt{DYNAMITE} code. While the methodology is extensively detailed and validated in \citet{zhu2020}, we outline the key configurations specific to this work below.

\subsection{The gravitational potential}\label{sec:gravpot}
The total gravitational potential in our model is constructed from three components: the stars, dark matter halo, and a central black hole. We note that our gravitational potential model explicitly neglects the gas components. Given that NGC~3957 is an S0 with low star formation rate ($\mathrm{SFR} = 0.26\, {\rm M}_\odot \mathrm{yr}^{-1}$), its molecular gas mass is estimated to be $M_{\mathrm{H}_2} \sim 5 \times 10^8\, {\rm M}_\odot$ via standard scaling relations with an average depletion time $t_{\rm dep} \sim 1.67$ Gyr \citep[typically 1-2 Gyr, e.g.,][]{Kennicutt2012, Boselli2014}. Furthermore, extensive surveys demonstrate that the neutral atomic gas ($M_{\mathrm{HI}}$) content in early-type galaxies is systematically suppressed, typically contributing $\lesssim 1-3\%$ of the stellar mass \citep[e.g.][]{serra2012,beasley2015}. Summing the molecular and atomic components, the total cold gas mass remains dynamically negligible, and its omission does not affect our derived orbital structures.

To construct the stellar mass distribution, we use the MGE fitted to the $3.6\mu$m Spitzer IRAC image. Again, such a choice is because the $r$-band image is strongly affected by the dust, but the $3.6\mu$m image is mildly affected and serves as a reliable tracer of the underlying stellar mass distribution. 

After the MGE fitting, we de-project the 2D Gaussian components into a 3D luminosity density by assuming a specific set of viewing angles ($\theta$, $\psi$, $\phi$). Here, $\theta$ and $\phi$ determine the line-of-sight orientation relative to the galaxy's principal axes, while $\psi$ is chosen to specify the rotation of the galaxy around the line-of-sight in the projected sky-plane. 
The resulting 3D luminosity density is then multiplied by a constant stellar mass-to-light ratio ($M_{*}/L$) to obtain the 3D mass density, from which the stellar gravitational potential is computed using the classical Chandrasekhar formula \citep{vandenbosch2008}.

Rather than fitting the three viewing angles ($\theta$, $\psi$, $\phi$) directly, we explore the intrinsic shape parameter space defined by axis ratios $p=Y/X$, $q=Z/X$, and the compression factor $u=X'/X$ \citep{vandenbosch2008}, where $X$, $Y$, $Z$ are the intrinsic long, intermediate, and short axis of the galaxy and $X'$ is the projected major axis. The conversion of the two sets of parameters follows Eq.(10) in \cite{vandenbosch2008}. This approach is computationally more efficient. For example, the deprojection of an axisymmetric system would have no constraint on the parameter $\phi$ but have a finite axis-ratio between $Y$ and $X$. In our modelling, we follow the approach by \cite{vandenbosch2008} and fix $u=0.999$ to allow for a slight degree of triaxiality of the galaxy.

For the dark matter halo, we adopt a spherical Navarro-Frenk-White \citep[NFW;][]{navarro1996} profile governed by two free parameters: dark matter virial mass $M_{200}$ and concentration $C$.
The potential also includes a central black hole characterised by a Plummer potential \citep{vandenbosch2008} with a fixed scale length of $a=0.001$~arcsec. The black hole mass $M_\bullet$ is fixed to be $10^{5.5} \rm{M}_\odot$. This assumption is safe since the sphere of influence of the central black hole is not resolved and would have negligible impact. We cannot directly constrain the black hole mass.

Overall, our gravitational potential is determined by five free hyper-parameters: the stellar mass-to-light ratio $M_{*}/L$, two parameters on the intrinsic shape of stellar distribution $p$, $q$, the dark matter virial mass $M_{200}$, and the concentration $C$.

\subsection{The orbit library}
For each model, with a set of hyper-parameters, we generate an orbit library containing tens of thousands of orbits. The orbits sampling follows the way described in \cite{vandenbosch2008}.

We first sample regular orbits according to a separable triaxial potential. The orbits are sampled from the three integrals of motion: energy $E$, second integral of motion $I_2$, and third integral of motion $I_3$ \citep{Binney2008}.
We sample two sets of $45 \times 25 \times 13 $ combinations of $(E,I_2,I_3)$ as initial conditions which include co-rotating and counter-rotating orbits. 

Box orbits are crucial for supporting the triaxial structures, we sample another set of box orbits by constructing the initial conditions on equipotential surfaces with the energy $E$, two spherical angles $\theta$ and $\phi$, which gives another set of $45 \times 25\times 13 $ orbits. 

It is worth noting that we employ a significantly denser orbit sampling compared to previous studies \citep[e.g.,][]{vandenbosch2008, zhu2018, jin2019, Ding2023}. For comparison, the CALIFA models used $21 \times 10 \times 7$ orbits \citep{zhu2018}, and the Fornax3D models used $55 \times 11 \times 11$ \citep{Ding2023}. The GECKOS data possess a much larger spatial extent and higher spatial resolution, thereby demanding greater degrees of freedom in the orbital library to achieve an accurate fit. Finally, to reduce numerical Poisson noise, every sampled orbit is dithered into $5^3$ sub-orbits by slightly perturbing the initial conditions.

\subsection{Fitting stellar luminosity density and kinematics}
We fit the luminosity density and stellar kinematics of the galaxy by weighting the orbits. 
The fitting constraints include the 2D surface brightness, the de-projected 3D luminosity density, and the 2D kinematic maps. 
Crucially, while the gravitational potential is derived from the $3.6\mu$m image, the luminosity constraints here are based on the dust-corrected $r$-band MGE (see Section \ref{sec:MGE}). This ensures that the orbital weights correctly reproduce the light distribution from which the MUSE kinematics were actually extracted.

The kinematic maps include the line-of-sight mean velocity $V$,
velocity dispersion $\sigma$, Gauss-Hermite coefficients $h_3$ and $h_4$. It should be noted that we do not directly fit $V$ and $\sigma$ maps, but the Gauss-Hermite coefficients $h_1$, $h_2$, $h_3$, and $h_4$. We extract similar luminosity and kinematic maps from the model superposed by orbits, then we obtain the solution of the orbit weights by minimizing the $\chi^2$ between the data and the model using a non-negative least squares (NNLS) method \citep{Lawson1974,lawson1995}.

We note that we do not mask the prominent dust lane when fitting the stellar kinematics. While dust extinction primarily biases the observed kinematics towards the near-side stars, our tests confirmed that masking the dust lane yields negligible differences in the recovered orbital phase-space and the global parameter constraints for NGC~3957 (see Appendix~\ref{sec:appendix_dust} for details).

\begin{figure*}
    \centerline{
        \includegraphics[width=2.1\columnwidth]{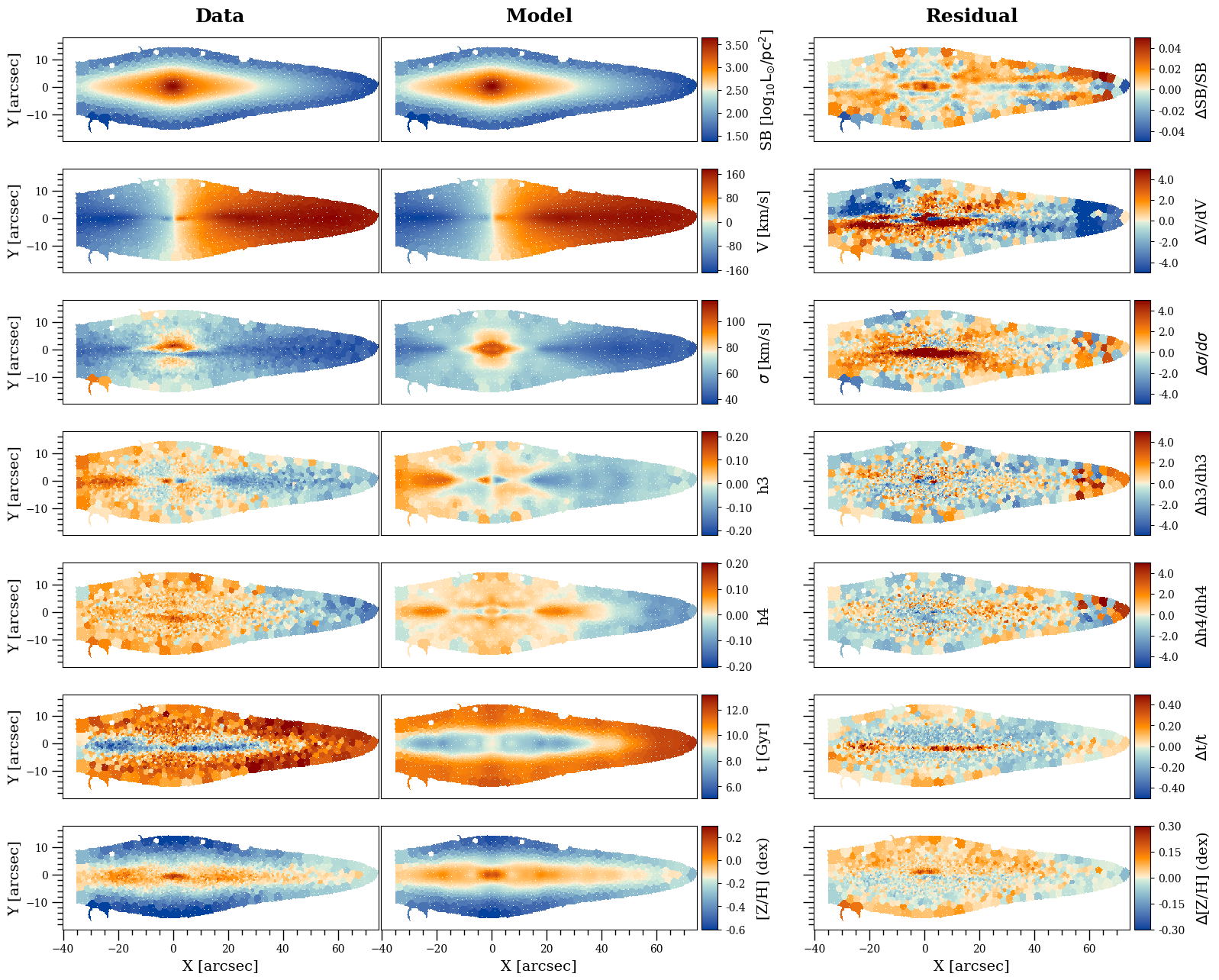}}
    \caption{Best-fitting population-orbit superposition model of
        NGC\,3957. {\em From left to right:\/} maps of the data, model, and
        residuals (model$-$data). {\em From top to bottom:\/} maps of the
        surface brightness SB, mean velocity $V$, velocity dispersion
        $\sigma$, Gauss-Hermite coefficients $h_3$ and $h_4$, light-weighted
        age $t$ and metallicity $[Z/H]$ of the stars. In the residual column, the kinematic maps are normalized by their respective observational uncertainties ($dV$, $d\sigma$, $dh_3$, and $dh_4$). For the surface brightness and age, the fractional residuals ($\Delta\text{SB}/\text{SB}$ and $\Delta t/t$) are shown, while the metallicity residual is given as the absolute difference ($\Delta[Z/H]$ in dex).
    }
    \label{fig:model_fitting}
\end{figure*}

We use an optimised grid-search algorithm \citep{zhu2018a} to identify the best-fitting hyper-parameters of the gravitational potential. We start with a model with an initial guess of the hyper-parameters, then we perform an iterative searching process with intervals of 0.05, 0.005, 0.005, 0.05, and 0.5 for the hyper-parameters $M_{*}/L$, $p$, $q$, $\log{M_{200}}$, and $C$, respectively. 
Among these models, we select the best-fitting models with $\chi^2-\chi^2_{\mathrm{min}}<200$, and sample new models around the selected ones. We continue the iterative process until an area of $\chi^2$ minimum is found and all models within a $3-\sigma$ confidence level around the minimum $\chi^2$ are calculated. 
At the end, one thousand models are calculated for the galaxy, and the parameter grid for NGC~3957 is presented in Fig.~\ref{fig:kinchi2_plot} in the Appendix. 
The model with the minimum $\chi^2$ is selected as the best-fitting model. As demonstrated in Fig.~\ref{fig:model_fitting}, we show the best-fitting model of NGC~3957, where the model matches the observed kinematic data in great detail. To account for modelling uncertainties, the $1\sigma$ confidence level is defined following \citet{Ding2023}:
\begin{align}\label{eqn:chi2}
\delta\chi^2 = \chi^2-\chi^{2}_{\mathrm{min}}<4\times\sqrt{2\times n_{\mathrm{GH}} \times N_{\mathrm{obs}}},
\end{align}
where $ n_{\mathrm{GH}}=4$ is the number of fitted kinematic moments, and $N_{\mathrm{obs}}=2229$ is the total number of Voronoi bins in our kinematic data.

\begin{figure*}
    \centerline{
        \includegraphics[width=2.2\columnwidth]{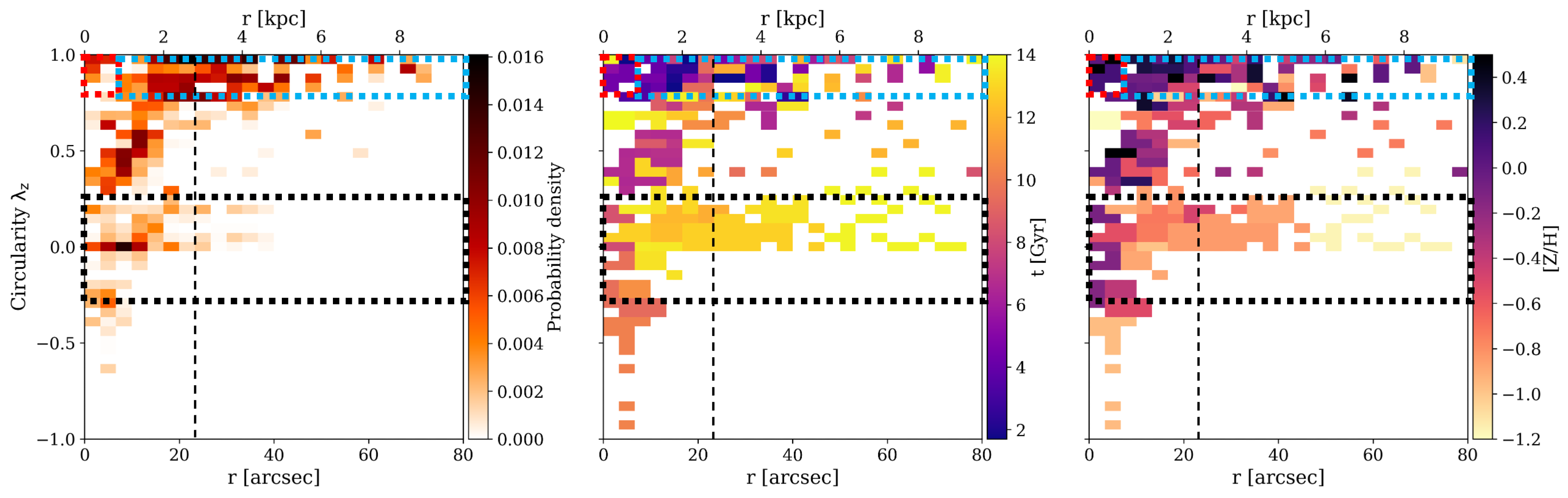}}
    \caption{Orbital decomposition of NGC\,3957.  The probability density
    distribution $p(r, \lambda_z)$ {\em (left panel)\/}, age distribution
    {\em (central panel)\/}, and metallicity distribution {\em (right panel)\/} of the stellar orbits in the phase space of
    time-averaged radius $r$ versus circularity $\lambda_z$. The
    probability densities are normalised to unity within the data
    coverage. All the distributions are averages of multiple best-fitting
    models that fall within the $1\sigma$ confidence level of the model
    hyper-parameters. 
    The coloured boxes with dashed line mark our orbit-based division into
    different components: 
    a dynamically cold NSD ($\lambda_z \geq0.8$ and $r<7$ arcsec, in red box)
    a dynamically cold main disc component ($\lambda_z \geq0.8$ and $r\geq7$ arcsec, in blue box) 
    and a dynamically hot component ($-0.25<\lambda_z <0.25$, in black box). 
    The black dashed vertical line represents the effective radius $R_{\rm e}=22.2$ arcsec.
    }
    \label{fig:phase_space_marked}
\end{figure*}

Once we fit the stellar kinematics, we extract the intrinsic orbital structure of the galaxy. We characterise each orbit using its time-averaged radius $r$ and its circularity $\lambda_z$. The circularity is defined as the orbital angular momentum around the short ($z$) axis normalized by the maximum angular momentum that is allowed by a circular orbit with the same binding energy. Thus, $\lambda_z \sim 1$ represents highly rotating short-axis tube orbits, and the $\lambda_z \sim 0$ represents mostly long-axis tube or box orbits. The resulting stellar orbit distribution for the best-fitting model of NGC~3957 is displayed in the left panel of Fig.~\ref{fig:phase_space_marked}.

\begin{figure*}
    \centerline{
        \includegraphics[width=2.\columnwidth]{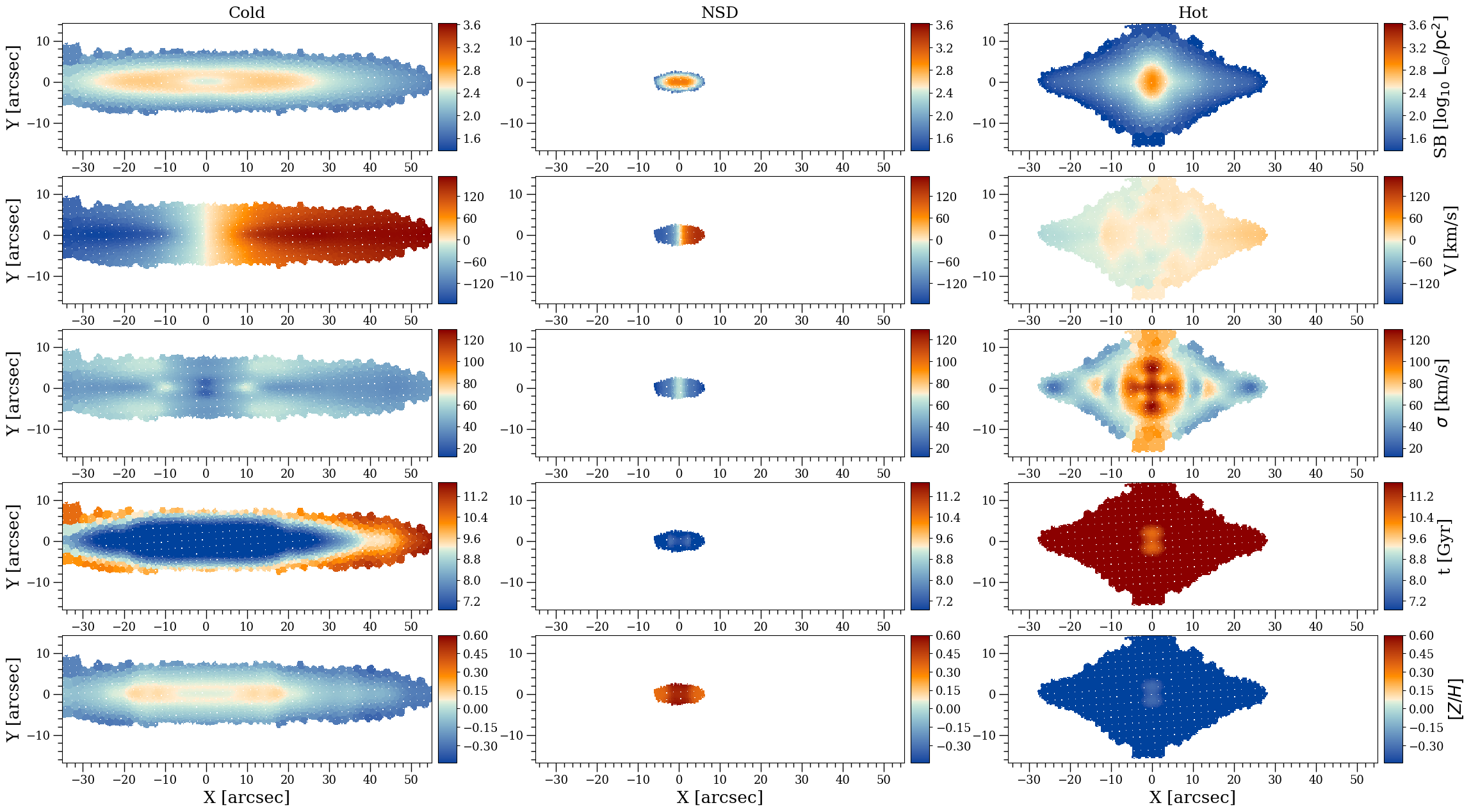}}
    \caption{Visualisation of the orbital decomposition of NGC~3957. 
    {\em From top to bottom:\/} maps of the surface brightness SB, mean velocity $V$, velocity dispersion $\sigma$, age $t$ , and metallicity $[Z/H]$.
    {\em From left to right:\/} maps of the the dynamically cold main disc component, NSD and the dynamically hot component.
    }
    \label{fig:structure_decmp_v2}
\end{figure*}

\subsection{The chemical tagging}
Following the kinematic fit, we proceed to assign age and metallicity values to the stellar orbits. We select all models that fall within the $1\sigma$ confidence interval determined by Eq.~(\ref{eqn:chi2}). 

To make the problem computationally tractable, we do not fit individual orbits. Instead, for each model, we group the orbits into approximately 100 orbital bundles. This is achieved by applying a 2D Voronoi binning scheme in the ($\lambda_z, r$) phase-space, ensuring that each resulting bundle carries a minimum orbital weight of 0.005. 

We assume that orbits within the same bundle $k$ share a single, uniform stellar age $t_k$ and logarithmic metallicity $[Z/H]_k$. The model-predicted age $t^i_{\mathrm{model}}$ and metallicity $[Z/H]^i_{\mathrm{model}}$ at any observational aperture $i$ are formulated as the luminosity-weighted average of all bundles contributing to that aperture:
\begin{align}
t^i_{\mathrm{obs}} &= \frac{\sum^{N_{\mathrm{bundle}}}_{k=1} t_k f^i_k}{\sum_k f^i_k}, \\
[Z/H]^i_{\mathrm{obs}} &= \frac{\sum^{N_{\mathrm{bundle}}}_{k=1} [Z/H]_k f^i_k}{\sum_k f^i_k}, \quad k=1,...,N_{\mathrm{bundle}},
\end{align}
where $N_{\mathrm{bundle}}$ is the total number of orbital bundles and $f^i_k$ is the luminosity contribution of bundle $k$ at aperture $i$. We note that we fit the stellar age on a linear scale, whereas the metallicity is fitted on a logarithmic scale ($[Z/H]$) to ensure consistency with the full-spectral fitting.

We apply a Bayesian statistical analysis using the No-U-Turn Sampler (NUTS) algorithm implementation in the Python package \texttt{pymc} \citep{Salvatier2016}. We first fit the age maps to solve for $t_k$, and subsequently fit the metallicity maps to find $[Z/H]_k$.
According to the Bayes theorem, deriving the posterior probability requires defining a prior and a data likelihood. 

Crucially, the central region ($r \le 18$~arcsec) of NGC~3957 is heavily obscured by the prominent equatorial dust lane, which introduces noticeable asymmetries in the age maps. To mitigate the resulting degeneracies in the chemical tagging, we introduce a physically motivated piecewise prior for the age fitting in this central region. Based on the expectation that dynamically colder, highly rotating orbits host younger stellar populations \citep[e.g.,][]{trayford2019}, we apply a circularity-dependent prior. Specifically, for orbits within $r \le 18$~arcsec and $\lambda_z \ge 0$, the prior mean is assigned following a empirical linear relation $\mu(t_k) = 12 \times (1 - \lambda_z)$~Gyr. For the outer regions ($r > 18$~arcsec) as well as all counter-rotating orbits ($\lambda_z < 0$), we retain the standard uniform prior. The mean value $\mu(t_k)$ is randomly sampled around this prior with a sampling width of five times of the standard deviation of the observed $t^i_{\rm obs}$ ($\sim 8$ Gyr).
The whole approach could effectively stabilise the age recovery against dust-induced noise in the central region.
We adopt a student’s t-distribution for the data likelihood \citep{Salvatier2016}, which allows for some outliers in the data and results in a robust fitting. For each $t_k$,
we run a chain with 2000 steps and take the last 500 steps to calculate the mean and $1\sigma$ values of $t_k$.

A similar Bayesian framework is then applied to the metallicity map. For $[Z/H]_k$, we adopt a uniform prior for $[Z/H]_k$ everywhere. A student’s t-distribution is used for the data likelihood in all cases to allow for outliers and ensure robust fitting. We run a chain with 2000 steps and take the last 500 steps to calculate the mean and $1\sigma$ values of $Z_k$.
The final best-fitting population-orbit model for NGC~3957 is presented in Fig.~\ref{fig:model_fitting}. The resulting model successfully reproduces the surface brightness, stellar kinematics, age, and metallicity maps simultaneously. We note that while the prominent equatorial dust lane leaves clear signatures in the observed kinematic maps (particularly in $\sigma$ and $h_4$), our model does not attempt to reproduce these highly dust-induced features.

\subsection{Orbital decomposition}\label{sec:decomposition}
In Fig.~\ref{fig:phase_space_marked}, we present the probability density, age, and metallicity distributions of the orbits from the best-fitting model of NGC~3957 in the phase space of $\lambda_z$ versus $r$.
Based on the stellar orbit distribution, we can decompose the galaxy into different dynamical components. 
While orbits with $\lambda_z > 0.8$ are generally considered dynamically cold, the presence of a distinct inner kinematic structure in the rotation curve (associated with the NSD; see Fig.~15 and Fig.~16 in \citealp{Fraser-McKelvie2025}) allows us to physically separate the cold orbits into two distinct parts at a radius of $\sim 7$~arcsec ($\sim 0.8$~kpc). 

Therefore, here we identify the three dynamical components: (1) a central NSD consisting of orbits with $\lambda_z \ge 0.8$ and $r \le 7$~arcsec; (2) a main disc consisting of orbits with $\lambda_z \ge 0.8$ and $r > 7$~arcsec; and (3) a dynamically hot component defined by orbits with $-0.25 < \lambda_z < 0.25$.

To better demonstrate the properties of each galaxy component, we show the maps of the surface brightness, mean velocity $V$, velocity dispersion $\sigma$, age $t$, and metallicity $[Z/H]$ for each component of NGC~3957 in Fig.~\ref{fig:structure_decmp_v2}. We see a fast-rotating, young, and spatial extended dynamically-cold disc in the left column, a fast-rotating old, and metal-rich central concentrated NSD in the middle column, and a non-rotating old, and metal-poor spheroidal dynamically-hot component in the right column.

\subsection{Modelling the bar-related structures}\label{sec:bar_modelling}
NGC~3957 is classified as a barred galaxy. However, we note that our adopted gravitational potential (with $u=0.999$, allowing for only a slight degree of triaxiality) does not explicitly include strongly non-axisymmetric bar orbits (e.g., $x_1$ orbits elongated along the bar and $x_2$ orbits oriented perpendicular to the bar) in the orbit library. 

Despite this limitation, previous studies have demonstrated that such nearly axisymmetric Schwarzschild models are still highly capable of capturing the phase-space signatures of a bar \citep[e.g.,][]{zhu2018a}. To reproduce the complex kinematics (such as the high velocity dispersion) induced by the unmodelled non-circular motions of the bar, the model dynamically compensates by populating dynamically ``warm'' orbits ($\lambda_z \sim 0.25-0.8$). 

To further validate the robustness of this approach in resolving central substructures, we performed a test using mock data from an isolated, barred RAMSES simulation (Fragkoudi \& Bieri, in prep) that self-consistently forms a bar and a NSD, without a central bulge \citep[see Appendix~\ref{sec:appendix_ramses} and Appendix C of][]{Camila2023}. As detailed in Appendix~\ref{sec:appendix_ramses}, when applying our modelling pipeline (without age and metallicity fitting) to this mock galaxy, we successfully recovered the broad orbital structures in the phase space. The model robustly captures both the dynamically warm orbits indicative of the bar, the highly concentrated cold orbits ($\lambda_z \ge 0.8$) characteristic of the NSD and main disc as well as the dynamically hot component ($|\lambda_z| \leq 0.25$). Therefore, to avoid over-interpreting these compensating warm orbits, we refrain from discussing the intrinsic properties of the stellar bar itself. Instead, we safely proceed to focus our analysis exclusively on the three components: the NSD, the dynamically cold main disc, and the dynamically hot component.

\section{Results}\label{sec:result}
In the last section, we established the orbital-superposition model for NGC~3957. We identify three dynamical component: a dynamically-cold main disc, a NSD, and a dynamically-hot component. In this section, we first discuss the general properties of the entire galaxy, then
visualise the three galaxy components and further look at the surface brightness, age, and metallicity radial profile of each component, respectively.

\subsection{General properties}
\begin{figure}
    \centerline{
        \includegraphics[width=1.\columnwidth]{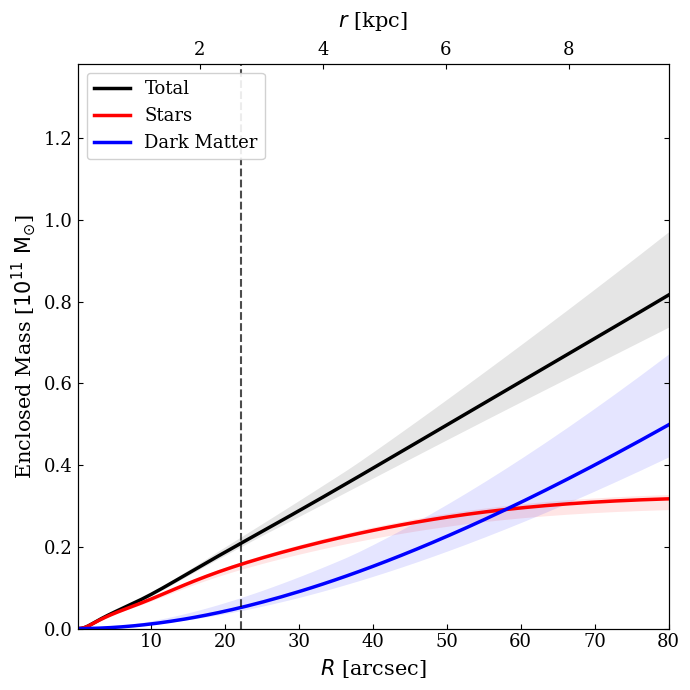}}
    \caption{
    Enclosed mass profiles for NGC~3957. The black, red, and blue lines represent the enclosed mass profiles of the total galaxy, stars, and the dark matter. The vertical dash line shows the effective radius $R_{\rm e}=22.2$ arcsec.
    }
    \label{fig:enclosedmassm_linear}
\end{figure}

Fig.~\ref{fig:enclosedmassm_linear} shows the enclosed mass profiles of the total, the stars, and the dark matter for NGC~3957. Within $R<58$ arcsec ($\sim 7.0$~kpc), the enclosed mass of stars is higher than that of the dark matter, beyond $R\sim60$ arcsec ($\sim 7.0$~kpc), the enclosed mass of dark matter becomes dominant. The enclosed mass of dark matter shows a rapid increase in the whole region, while the enclosed mass of stars increases strongly within $R<50$ arcsec ($\sim 6.0$~kpc) and becomes flat in the outer region. The dark matter mass fraction is $f_{\rm DM,e} = 28 \pm 7 \%$ within the effective radius ($R_{\rm e} = 22.2$~arcsec or $\sim 2.7$~kpc) and $f_{\rm DM,2e} = 45 \pm 10 \%$ within $2R_{\rm e}$.

We define the luminosity fraction of each component as:
\begin{align}
f =\sum_{k}^{\mathbb{C}} f_k,
\end{align}
where $f_k$ is the luminosity fraction of orbital bundle $k$ that belongs to the region $\mathbb{C}$ in the $\lambda_z-r$ phase-space. For example, $\mathbb{C}$ is $-0.25<\lambda_z<0.25$ for the dynamically-hot component. We note that the total luminosity fractions are calculated within the limits of our data coverage ($R < 80$~arcsec, corresponding to $\sim 9.6$~kpc).

Similarly, we can
define the luminosity-weighted mean value of age and metallicity of each component by
\begin{align}
\left\langle t \right\rangle =& \frac{1}{f} \sum_{k}^{\mathbb{C}} t_k f_k,\\
\left\langle [Z/H] \right\rangle =&\frac{1}{f} \sum_{k}^{\mathbb{C}} [Z/H]_k f_k,
\end{align}
where $t_k$ and $[Z/H]_k$ are the age and logarithmic metallicity of the orbital bundle $k$ residing in the phase-space region $\mathbb{C}$.

\begin{figure*}
    \centerline{
        \includegraphics[width=1.5\columnwidth]{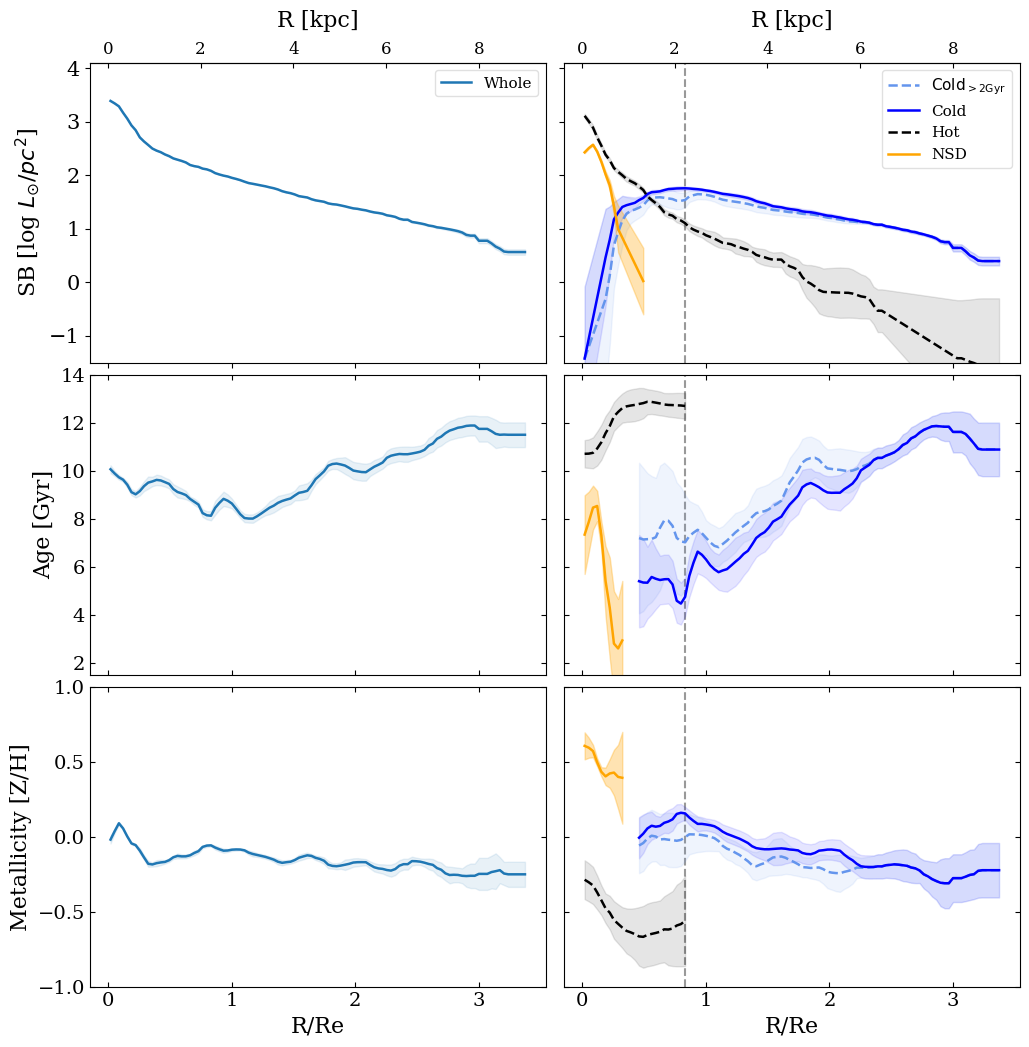}}
    \caption{Surface brightness, age, and metallicity radial profiles of the whole galaxy, the dynamically cold main disc, NSD and the dynamically hot component for NGC~3957. We show the profiles of the whole galaxy ({\em left panels\/}) with blue solid curve, and the profiles of dynamically cold disc, NSD and dynamically hot component ({\em right panels\/}) with blue solid curve, orange solid curve and black dashed curve. The shadowed areas indicate the scatter of the profiles of models that fall within the 1$\sigma$ confidence level. To investigate the origin of the age gradient, we also show the profiles for the cold disc calculated by excluding stars younger than 2 Gyr (represented by the blue dashed lines in the right panels). The profiles of three components are shown in the regions where they contribute at least 10\% of the total surface brightness. The dashed vertical line represents the bar radius $R_{\rm bar}=18.5$ arcsec. 
    }
    \label{fig:profiles}
\end{figure*}

\begin{figure*}
    \centerline{
        \includegraphics[width=2.1\columnwidth]{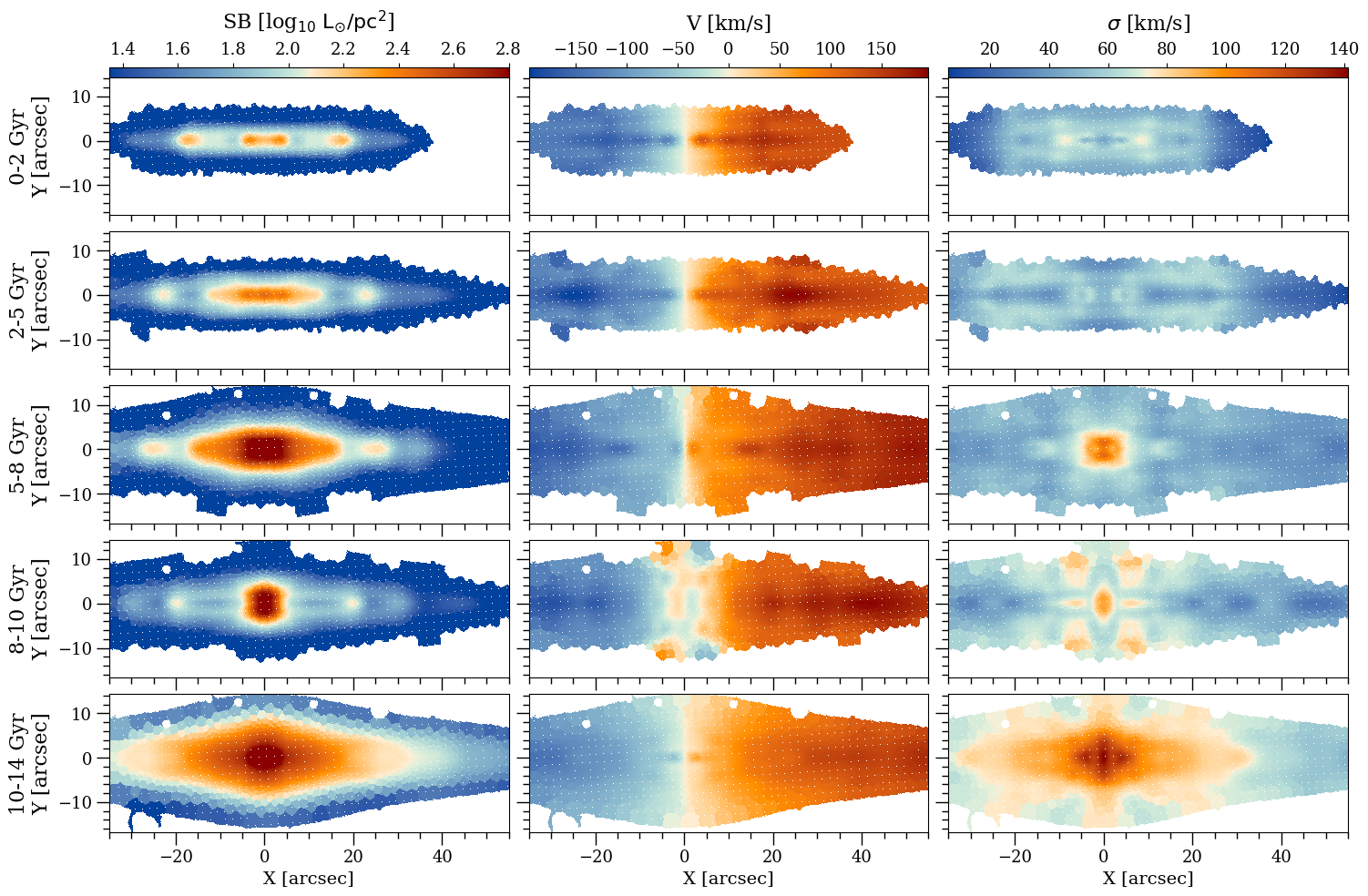}}
    \caption{Intrinsic property maps of stellar populations in NGC~3957 decomposed into five age bins: \(0-2\), \(2-5\), \(5-8\), \(8-10\), and \(10-14\) Gyr (from top to bottom). For each age bin, we show (from left to right): surface brightness (SB), mean line-of-sight velocity ($V$), and velocity dispersion ($\sigma$). The maps are reconstructed from the best-fitting orbit-superposition model.
    }
    \label{fig:Age_bins_whole}
\end{figure*}

\begin{figure*}
    \centerline{
        \includegraphics[width=2.1\columnwidth]{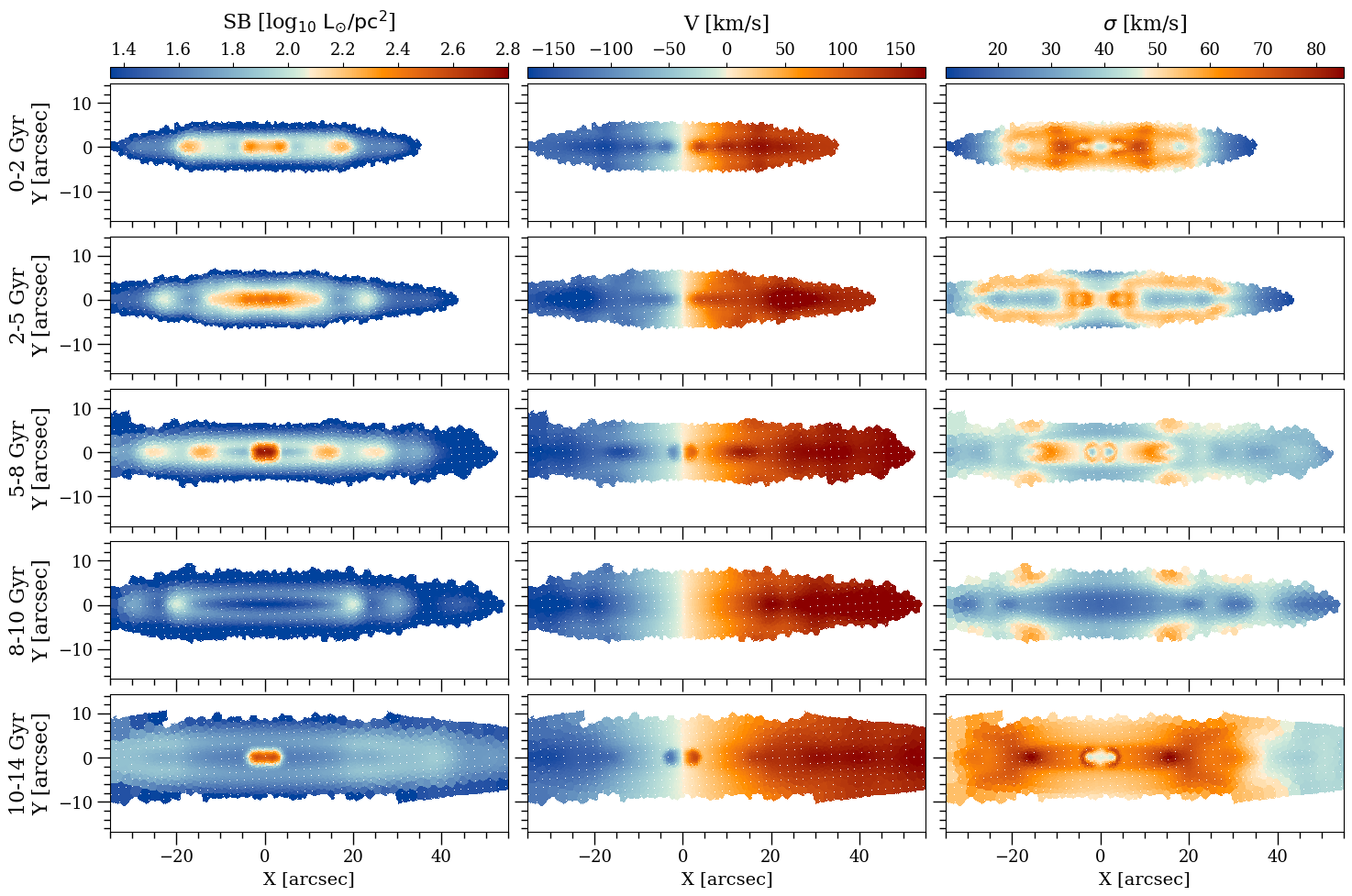}}
    \caption{Same as Fig.~\ref{fig:Age_bins_whole}, but we only select the disc stars ($\lambda_z>0.8$).
    }
    \label{fig:Age_bins_disk}
\end{figure*}

\subsection{Results of the orbital decomposition}\label{sec:result_decmp}
In the left column of Fig.~\ref{fig:structure_decmp_v2}, we show the dynamically-cold main disc. It has a stellar mass $M_{\star}= (1.38\pm0.15) \times 10^{10} {\rm M}_\odot$. The luminosity fraction of this component is $f_{\rm main\ disc}=0.45 \pm 0.02$. 
To quantify its spatial extent, we perform a 2D double-exponential fit to its surface brightness map. The best-fit yields a radial scale length of $h_R = 39.87 \pm 4.02$~arcsec ($4.79 \pm 0.48$~kpc) and a vertical scale height of $h_z = 4.22 \pm 0.24$~arcsec ($0.51 \pm 0.03$~kpc). The resulting ratio of $h_R/h_z \approx 9.4$ provides morphological evidence that this dynamically-selected component might correspond to a thin disc. 

The main disc is characterised by a fast-rotating pattern, reaching $V \sim 180$~km/s beyond $r \gtrsim 20$~arcsec, while maintaining a very low overall velocity dispersion of $\sigma \sim 30-60$~km/s. The overall average age of the main disc is $t = 7.7\pm0.5$~Gyr. As shown in the age map, it exhibits a clear vertical age gradient: younger stars are strongly concentrated towards the midplane $z\lesssim5$ arcsec, while the stellar age sharply increases at larger vertical heights. The metallicity of the main disc is $[Z/H] = -0.02\pm 0.04$~dex, though the metallicity decreases slightly towards the outer edges.

The middle column of Fig.~\ref{fig:structure_decmp_v2} presents the NSD. It has a stellar mass $M_{\star}=(8.5 \pm 1.0) \times 10^9 {\rm M}_\odot$. The luminosity fraction of this component is $f_{\rm NSD}=0.03 \pm 0.01$. The NSD is a highly compact structure concentrated within the central $\sim 0.8$~kpc. It exhibits extremely rapid rotation (maximum $V\sim 180$~km/s) and a characteristically low velocity dispersion ($\sigma \sim 20-60$~km/s). Consistent with our physically motivated prior, the NSD is found to be a relatively young structure, with an overall average age of $t =6.9\pm0.4$~Gyr. It is also the most metal-rich component in the galaxy, with an average metallicity of $[Z/H] =0.49 \pm 0.06$~dex.

The right column of Fig.~\ref{fig:structure_decmp_v2} shows the dynamically-hot component. It has a stellar mass $M_{\star}=(4.4\pm0.8) \times 10^9 {\rm M}_\odot$. The luminosity fraction of this component is $f_{\rm hot}=0.14\pm0.02$. It shows a spheroidal shape, which extends to $r\sim 30$ arcsec. To characterize its surface brightness distribution, we perform a 2D elliptical Sérsic fit. We obtain a best-fitting Sérsic index of $n=3.00 \pm 0.51$ and an effective radius (measured along the semi-major axis) of $R_{\rm hot} = 18.2 \pm 4.3$~arcsec ($2.2 \pm 0.5$~kpc) with an axis ratio of 0.6.
The hot component shows very weak rotation ($V \sim 20$~km/s) and a high, centrally-peaked velocity dispersion ($\sigma \sim 130$~km/s at the centre, decreasing outwards). 
The hot component is globally ancient, with an overall average age of $t=11.9\pm0.6$ Gyr. Its metallicity is generally lower than the disc components, showing an average value of $[Z/H] = -0.39 \pm 0.20$~dex. 

To contextualise the physical scales of these components, we compare them with the galactic bar. Following the kinematic approach introduced by \citet{jin2025b}, we estimate the bar radius $R_{\rm bar}$ as the boundary where dynamically cold orbits ($\lambda_z \ge 0.8$) begin to dominate the whole galaxy. By computing the average cold orbit luminosity fraction $f_{\rm cold}$ within a moving window of $0.42$~arcsec and a step size of $1$~arcsec, we identify $R_{\rm bar}$ as the innermost radius where $f_{\rm cold} \ge 0.5$. This yields $R_{\rm bar} = 18.5$~arcsec ($2.22$~kpc) for our best-fitting model. Although our Schwarzschild model lacks an explicit non-axisymmetric bar potential, our mock tests confirm that this kinematic threshold still provides a highly reliable approximation of the true bar extent (see Appendix~\ref{sec:appendix_ramses}). 

Comparing the structural sizes, we find a coherent physical picture. 
First, the effective radius of the dynamically-hot component ($R_{\rm hot} = 18.2$~arcsec or $2.18$~kpc) matches the bar radius perfectly. This coincidence suggests that the hot component may partially represent the dynamically hot orbits of the stellar bar (i.e., the vertically thickened B/P structure). This interpretation is supported by our mock test on a simulated bulgeless galaxy (Appendix~\ref{sec:appendix_ramses}), which demonstrates that highly eccentric bar orbits could naturally contaminate the low-circularity parameter space. In light of this, the 14\% mass/luminosity fraction (assuming a constant $M_{*}/L$)  measured for the hot component here is best interpreted as an upper limit for the true fraction of the central bulge.
Second, the compact NSD is deeply embedded within the very centre of the bar. Finally, the scale length of the main cold disc ($h_R = 39.87$~arcsec or $4.79$~kpc) is roughly twice the size of $R_{\rm bar}$, indicating that the cold disc extends far beyond the bar-dominated region.

\subsection{Surface brightness, age, and metallicity radial profiles}\label{sec:profile}

To further quantify the spatial variations, we reconstruct the 3D luminosity, age and metallicity distributions of each orbital component. 
We project these components face-on and extract the radial profiles from the projected maps. We compute these profiles for all models falling within the $1\sigma$ confidence level. We take the average of these profiles from models within the $1\sigma$ confidence level as the mean profile, and their scatter as the $1\sigma$ uncertainty.

In Fig.~\ref{fig:profiles}, we show the radial profiles of surface brightness, age and metallicity for the whole galaxy, the dynamically-cold main disc, NSD, and hot component. As demonstrated in \citet{zhu2020}, stellar populations cannot be reliably recovered at radii where a component contributes $\leq 10\%$ to the total luminosity. Therefore, we only display the profiles within regions where each component meets this surface brightness threshold.

The age of the whole galaxy shows a small decrease from $R=0$ to $R\sim 1R_{\rm e}$ and increases outwards. The metallicity profile of the whole galaxy is overall flat except for a small increase in the  inner $r\sim 0.4R_{\rm e}$ region.

The dynamically cold main disc presents a significant positive age gradient, rising from $\sim 5$~Gyr in the inner $0.8R_{\rm e}$ to over $\sim 12$~Gyr in the outer disc ($R \sim 3 R_{\rm e}$). The age difference of $\sim 7$~Gyr between the inner and outer regions indicates an ``outside-in'' quenching scenario, where star formation ceased in the outskirts long ago but was sustained in the inner regions. We observe a flat age gradient within the bar radius.
To determine whether the positive age gradient in the main disc is merely a recent effect driven by residual star formation or a long-term evolutionary feature, we perform a test by excluding all stars younger than 2 Gyr from the main disc profile (light-blue dashed line in Fig.~\ref{fig:profiles}). While phase-space maps (Fig.~\ref{fig:phase_space_marked}) show that young stars are distributed everywhere across the main disc (with a higher concentration in the inner $\sim20$ arcsec), excluding them does not erase the gradient. Instead, the age of the inner disc ($\sim 0.5-1R_{\rm e}$) shifts upwards to $\sim 7$~Gyr, but the overall positive slope from the inner to the outer disc remains robust and significant. This confirms that the positive age gradient is not a transient feature caused by recent star formation, but rather a long-term evolutionary imprint.

The metallicity radial profile of the main disc shows a slight negative gradient from the inner $0.8R_{\rm e}$ to outer regions. We observe a positive gradient within the bar radius. After excluding the young stars, the metallicity radial profile is also very similar to the one of all stars in the main disc.

The NSD is identified as the youngest and most metal-rich structure, with a strong negative age gradient. Its stellar age decreases from 8 Gyr at the centre to 3 Gyr in the outer region. It displays a slight negative metallicity gradient, dropping from $[Z/H] \sim 0.6$~dex at the very centre to $\sim 0.4$~dex at its outer edge. 

The dynamically hot component is significantly older and more metal-poor than other components. It displays a positive age gradient, rising from 10 Gyr in its inner region to 13 Gyr in the outer region. It shows a negative metallicity gradient, dropping from $[Z/H] \sim -0.2$~dex at the very centre to $\sim -0.6$~dex at its outer edge.

\subsection{Assembly of NGC~3957}\label{sec:assembly}

In this section, we investigate the assembly history of NGC~3957 by dissecting the galaxy into different stellar age populations. Based on the orbit superposition model, we reconstruct the intrinsic 2D maps of surface brightness SB, velocity $V$, and velocity dispersion $\sigma$ for stars within five different age bins: \(0-2\), \(2-5\), \(5-8\), \(8-10\), and \(10-14\) Gyr. To understand the formation of the disc specifically, we also analyse the subset of stars with circularity $\lambda_z > 0.8$ (hereafter "disc stars") for each epoch. The maps for the whole galaxy and the disc stars are shown in Fig.~\ref{fig:Age_bins_whole} and Fig.~\ref{fig:Age_bins_disk}, respectively.

The age bin \(10-14\) Gyr represents the oldest stars in NGC~3957. Globally, they form a massive, extended structure with significant rotation ($V \sim 60$ km/s) and relatively high dispersion ($\sigma \gtrsim 100$ km/s) in the centre. When focusing on the disc stars, we find they dominate the central region with high surface brightness, exhibiting a strong rotating structure likely associated with the early formation of the NSD.

The age bin \(8-10\) Gyr reveals a striking kinematic decoupling. While the outer parts of the galaxy show prograde rotation, the very centre ($r \lesssim5$ arcsec) displays a weak counter-rotating core. For the disc component ($\lambda_z > 0.8$), this epoch appears to be a relatively quiescent period. The surface brightness is generally low and diffuse, lacking a strong central concentration. 

At the epoch of \(5-8\) Gyr, the galaxy shows a more regular, prograde-rotating structure. Star formation activity in the disc appears to reignite, forming a more defined structure with increasing rotation and moderate velocity dispersion ($\sigma \sim 40-50$ km/s). Crucially, this is the epoch where the NSD becomes highly prominent. The cold disc maps (Fig.~\ref{fig:Age_bins_disk}) show a massive concentration of bright, fast-rotating stars in the central $\sim 5$~arcsec, corresponding to the formation peak of the NSD.

At the epoch of \(2-5\)~Gyr, we observe a dramatic spatial shrinkage of the star-forming regions. Star formation begins to fade in the outskirts and becomes strongly confined to the inner regions ($R \lesssim 20$~arcsec), accompanied by a decreasing velocity dispersion. A similar inward-shrinking trend is clearly visible in the cold disc stars.

In the youngest age bin \(0-2\) Gyr, the stellar population resides in a very thin, dynamically cold disc structure ($\sigma \sim 20-60$~km/s for the whole galaxy). This recent star formation is strongly concentrated in the inner equatorial plane $R\lesssim20$~arcsec.

\section{Discussion}\label{sec:discussion}
\subsection{Morphology of the dynamical components}\label{sec:discussion_morphology}
As detailed in the Sect.~\ref{sec:result_decmp} regarding the spatial extents of our dynamically decomposed structures, we now compare our morphological parameters with previous 2D photometric decompositions in the literature \citep{pohlen2004}. 
For the central hot component, our dynamically derived effective radius ($R_{\rm hot} = 18.2$~arcsec or $2.18$~kpc) is larger than their purely photometric bulge estimate ($R_{\rm e} = 9.8$~arcsec or $1.18$~kpc). However, both methods yield a remarkably identical axis ratio of $q = 0.6$. Furthermore, the Sérsic indices are highly consistent: we obtain $n = 3.00$ from our dynamical hot component, which closely matches their photometric bulge index of $n = 3.3$. This strong morphological similarity suggests that despite the difference in effective radius, both approaches are broadly capturing the extended nature of the same large-scale central structure.

Regarding the main disc, the measurements also present notable differences, which naturally arise from the different foundational principles of the two methods. First, the vertical definitions differ mathematically. While we model the vertical profile using an exponential distribution defined by the scale height $h_z$, \citet{pohlen2004} adopted a $\text{sech}^2(z/z_0)$ vertical profile. Mathematically, the $\text{sech}^2$ function intrinsically yields a $z_0$ parameter that is roughly twice the equivalent exponential scale height ($z_0 \approx 2 h_z$). Bearing this mathematical relation in mind, we can directly compare the values. Our dynamically derived radial scale length is $h_R = 39.87$~arcsec ($4.79$~kpc) with a vertical scale height of $h_z = 4.22$~arcsec ($0.51$~kpc), yielding a geometric ratio of $h_R / h_z \approx 9.4$. In contrast, the photometric thin disc estimates are $h_R = 21.6$~arcsec ($2.59$~kpc) and $z_0 = 6.0$~arcsec ($0.72$~kpc). Converting their $z_0$ to an equivalent exponential scale height gives $h_{z, \rm eq} \approx 3.0$~arcsec ($0.36$~kpc), which results in a smaller ratio of $h_R / h_{z, \rm eq} \approx 7.2$. 

These discrepancies highlight the inherent differences between the techniques. Purely photometric models separate structures based solely on surface brightness profiles. They might underestimate the radial scale length $h_R$ because it is difficult to cleanly decouple the underlying thin disc from the overlapping, bright central bar and NSD. Conversely, our dynamical approach probably does better to isolate components based on orbital circularity $\lambda_z$. Finally, the treatments of the prominent equatorial dust lane are fundamentally different: while the previous photometric study opted to simply mask the dust-obscured regions (discarding crucial midplane information), our analysis utilises a dust-corrected MGE to physically reconstruct the intrinsic stellar mass distribution before any kinematic decomposition.

\subsection{The formation of the bar and the NSD}\label{sec:NSD}

\begin{figure}
    \centerline{
        \includegraphics[width=1.\columnwidth]{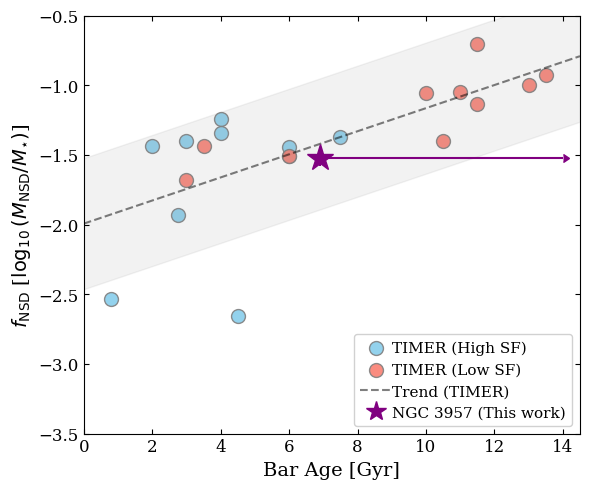}}
        \caption{
        The mass fraction of the NSD as a function of the age of the bar. We compare our results for NGC~3957 (purple symbol) to the values obtained by the TIMER survey \citep{Camila2025} for galaxies with a high (blue points) and low (red points) SFR. The shaded region shows the $1\sigma$ uncertainty of TIMER data points. For NGC~3957, the lower limit of the bar age is inferred from the age of the NSD. The rightward arrow indicates that the upper limit is unconstrained and is thus conservatively extended to 14 Gyr.
        }
    \label{fig:f_vs_age_NSD}
\end{figure}

Our dynamical decomposition reveals that the NSD in NGC~3957 is the youngest structural component in the galaxy, with a luminosity-weighted mean age of $6.9 \pm 0.4$~Gyr. Its  central ages reaching $\sim8$ Gyr and the outer edges decreases to $\sim3$ Gyr (Fig.~\ref{fig:profiles}). Theoretical and simulation-based works suggest that NSDs are formed by gas inflows driven by the bar potential, which funnel material toward the Inner Lindblad Resonance shortly after the bar itself becomes stable 
\citep{athanassoula1992,regan2004,athanassoula2005,emsellem2015,fragkoudi2016}. Consequently, the formation epoch of the NSD serves as a robust proxy for the formation epoch of the bar \citep{gadotti2015,Camila2023}. 
The relatively young age of the NSD ($\sim 7$~Gyr) implies that the bar in NGC~3957 might have formed at least 7 Gyr ago ($z \gtrsim 0.8$)
As bars originate from dynamical instabilities within a pre-existing cold stellar disc, its true formation age likely falls between the NSD formation ($\sim 7$~Gyr) and the age of the ancient stellar populations that comprise the early inner disc ($\sim 12-14$~Gyr). This indicates that the bar in NGC~3957 is a long-lived (at least $\sim7$~ Gyr ago), stable structure.

In Fig.~\ref{fig:f_vs_age_NSD}, we compare the mass fraction of the NSD ($f_{\rm NSD}=0.03$) and our inferred bar age range for NGC~3957 to the scaling relations obtained by the TIMER survey for face-on barred galaxies \citep{Camila2025}. In the TIMER sample, low-SFR galaxies generally host older bars (mostly $>9$~Gyr, with some between $6-9$~Gyr, and very few $\sim3$~Gyr). Our inferred bar age ($\ge 7$~Gyr) falls reasonably well within this expected range. Although our results are derived from orbit-superposition dynamical modelling, which is different from the TIMER results based on spectral synthesis of nuclear rings, our bar age estimation of NGC~3957 aligns well with the observed correlation trend. This provides an encouraging cross-validation for our model-derived component masses and ages.

A key feature of the NSD in our model is the distinct negative age gradient, where the stellar populations are oldest in the very centre ($\sim 8$~Gyr) and become younger at larger radii ($\sim 3$~Gyr; see Fig.~\ref{fig:profiles}). This matches the `inside-out' growth scenario for NSDs proposed by \citet{Bittner2020} and further confirmed by the detail star formation histories in \citet{Camila2023}. In the case of NGC~1433, \citet{Camila2025} found that while the core of the NSD formed $\sim 9.5$ Gyr ago, the outskirts were populated by younger stars formed in subsequent star formation episodes.

The presence of a relatively young NSD concentrated within the central $R < 7$~arcsec aligns with the scenario of prolonged, bar-driven secular evolution \citep[e.g.,][]{Camila2023}. The continuous funnelling of gas toward the central regions implies that as long as gas is available in the disc, new stars can be formed in the nuclear ring or within the NSD. Our dynamical model captures this relatively young component, providing a possible direct view of recent or ongoing bar-driven assembly in the central region. This suggests that even if the galaxy as a whole exhibits a low SFR, its core might have maintained star formation activity for an extended period due to the bar's influence.

Furthermore, our model indicates that the NSD is exceptionally metal-rich ($\langle [Z/H] \rangle \approx 0.49$~dex) in NGC~3957. This chemical enhancement can be naturally explained within the bar-driven inflow framework. As gas is funnelled inwards, it settles into the deep central potential well, where it can rapidly form stars and self-enrich \citep{ellison2011,Camila2023,Fraser-McKelvie2026}. This efficient self-enrichment process likely causes the resulting NSD to reach metallicities significantly higher than those of the main disc, a feature clearly recovered in our chemo-dynamical decomposition.

\subsection{The assembly history of NGC~3957 and the survival of its disc}\label{sec:assembly_and_NSD}

\begin{figure}
    \centerline{
        \includegraphics[width=1.\columnwidth]{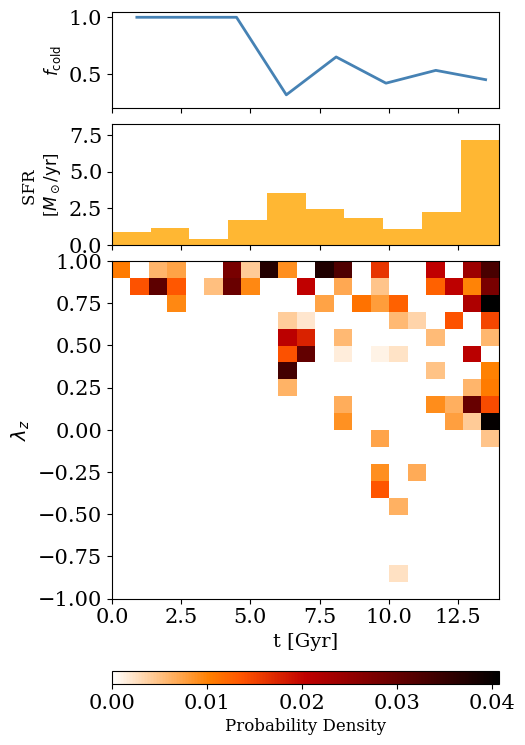}}
    \caption{
    Assembly history of NGC~3957. The bottom panel shows the probability density distribution $p(t, \lambda_z)$ of the stellar orbits in the phase space of age $t$ versus circularity $\lambda_z$, normalized to unity. The middle panel displays the star formation history (SFH), expressed as the absolute star formation rate (SFR) in ${\rm M}_\odot/\text{yr}$. The top panel illustrates the evolution of the cold disc fraction $f_{\text{cold}}$ as a function of stellar age.
    }
    \label{fig:Assembly}
\end{figure}

To constrain the assembly history of NGC~3957, we analyse the orbital distribution of its stellar populations in phase-space circularity $\lambda_z$ versus age $t$ (bottom panel of Fig.~\ref{fig:Assembly}), alongside the star formation history (SFH, middle panel) and the cold disc fraction evolution (top panel). Here, the cold disc fraction $f_{\rm cold}$ is evaluated for all radii with $\lambda_z \ge 0.8$, including both the NSD and the main disc. We apply a running window in the $t$ direction with a 1~Gyr width and a 1~Gyr step size to calculate the evolution of $f_{\rm cold}$. The absolute SFR is derived by multiplying the 1D stellar age distribution by the model-predicted total stellar mass ($3.14 \times 10^{10} {\rm M}_\odot$) within our data coverage ($R < 80$~arcsec). We note that this calculation provides the formation rate of the currently surviving stellar mass. It does not account for the evolutionary stellar mass loss (i.e., gas returned to the interstellar medium from dying stars), which can be up to $\sim 30\%-40\%$ for the oldest stellar populations assuming a Chabrier IMF \citep[e.g.,][]{Bruzual2003, Courteau2014}. Therefore, the SFR at older epochs ($t > 8$~Gyr) should be strictly interpreted as a lower limit to the true initial SFR.

The earliest epoch ($t > 12$~Gyr ago) is characterised by a broad distribution of orbits. A significant fraction of stars occupy dynamically hotter orbits ($\lambda_z \lesssim 0.5$), which likely represents the early assembly of the hot spheroid (also see Fig.~\ref{fig:Age_bins_whole}, bottom panel). We also find dynamically cold orbits ($\lambda_z \ge 0.8$). This early co-existence suggests that an initial disc was already forming alongside the hotter components, providing the necessary dynamical environment from which the stellar bar could later emerge. Following this initial build-up, the overall star formation appears to decline, reaching a minimum around $\sim 11$~Gyr ago.

A notable transition occurs $\sim 8-10$~Gyr ago. In the bottom panel of Fig.~\ref{fig:Assembly}, we observe the emergence of a counter-rotating stellar population ($\lambda_z \lesssim -0.25$), which is also visible in the corresponding age bin in Fig.~\ref{fig:Age_bins_whole}. 
Interestingly, the cold disc fraction ($f_{\rm cold}$) does not significantly collapse during this epoch (Fig.~\ref{fig:Assembly}, top panel) or later, indicating that the main prograde disc was not destroyed. Furthermore, the overall line-of-sight velocity dispersion appears relatively lower ($\lesssim 40$~km/s) compared to the other epoch (Fig.~\ref{fig:Age_bins_disk}). These clues suggest that this counter-rotating signature might be the result of a specific accretion event such as a minor merger and/or the accretion of misaligned gas from the environment \citep[e.g.][]{Davis2011,Algorry2014} rather than a violent major merger that would typically erase the cold disc signature.

This period is followed by a significant resurgence in star formation, peaking around 5-8~Gyr ago at a SFR $\approx 3.5~{\rm M}_\odot/\text{yr}$. It is plausible that existing and/or newly accreted gas (potentially linked to the aforementioned counter-rotating features at $\sim 10$~Gyr) was subjected to the non-axisymmetric torques of the pre-existing stellar bar. In the classic picture of bar-driven gas dynamics, gas loses angular momentum and streams inwards along highly elongated orbits (often visible as prominent dust lanes) before piling up to build a compact NSD or nuclear ring at the centre \citep[e.g.,][]{athanassoula1992, Sormani2020}.

Importantly, both high-resolution hydrodynamical simulations \citep[e.g.,][]{tress2020,Sormani2020} and recent integral-field surveys such as TIMER and PHANGS \citep[e.g.,][]{neumann2020, Fraser-McKelvie2020, Pessa2023} demonstrate that star formation is not strictly confined to the central nuclear ring. The gas compression and shocks occurring along the inflow streams can also trigger in-situ star formation directly within the bar itself. 

We note that our orbit-superposion model does not explicitly contain non-axisymmetric bar orbits (i.e., the $x_1$ and $x_2$ orbital families). However, as demonstrated in \citet{zhu2018a} and validated by our tests on a simulated barred galaxy (see Appendix~\ref{sec:appendix_ramses}), the model mimics the kinematic signature of a bar mostly by populating dynamically warm orbits ($\lambda_z \sim 0.25-0.5$). Therefore, the simultaneous emergence of relative young stars ($t \sim 5-8$~Gyr) within the cold component (mostly in central NSD $r \lesssim 5$~arcsec, see Fig.~\ref{fig:Age_bins_disk}) and across the dynamically warm component ($0.25<\lambda_z<0.8,\ r \lesssim R_{\rm bar}=18.5$~arcsec, see Fig.~\ref{fig:Age_bins_whole}) in the $\lambda_z-t$ space (Fig.~\ref{fig:Assembly}, middle panel) provides a compelling, integrated picture. It captures both the growth of the NSD at the centre and the concurrent star formation occurring along the bar as it actively funnels gas inwards.

After $\sim 5$~Gyr ago, the galaxy appears to settle into a relatively quiescent and secularly evolving phase. The SFR at $t=5$~Gyr is $1.66~{\rm M}_\odot/\text{yr}$ (specific star-formation rate, sSFR $=5.3 \times10^{-11}~\text{yr}^{-1}$) and reaches the minimum at $t\sim 3-4$~Gyr with SFR~$=0.39~{\rm M}_\odot/\text{yr}$ (sSFR $=1.2 \times10^{-11}~\text{yr}^{-1}$). The cold disc fraction remains high and stable (Fig.~\ref{fig:Assembly}, top panel), and star formation over the last 5 Gyr is largely confined to the dynamically cold orbits in the innermost regions ($r<20$~arcsec). 

This interpretation of a quiet late-time evolution is consistent with deep imaging of NGC~3957 from VLT/VIMOS \citep{jablonka2010} with detection limits $\mu_R = 30.6 ~{\rm mag}/{\rm arcsec}^2$ and $\mu_V = 31.4 ~{\rm mag}/{\rm arcsec}^2$, which revealed no prominent tidal streams, shells, or plumes that are typically associated with recent merger events. 
Furthermore, in the context of disc fragility, recent N-body simulations by \citet{Pablo2023} demonstrated that while NSDs are remarkably resilient to intermediate-mass-ratio (1:4) dry mergers, the surrounding kpc-scale thin discs are highly fragile. A single 1:4 merger event can dramatically thicken the main thin disc and erase its high-$\lambda_z$ kinematic signature. They found that the simultaneous existence of a well-defined, dynamically-cold main disc and a compact NSD strongly precludes any intermediate-mass merger events in the recent past ($t < 10$~Gyr). 
For NGC~3957, the sustained high cold disc fraction over the last 5 Gyr aligns with such finding, indicating NGC~3957 may not have any intermediate-mass merger events within, at least the recent $\sim5$~Gyrs.
Our orbital decomposition provides a kinematic perspective that is largely in agreement with both morphological evidence and the survival of dynamically cold disc, portraying a galaxy whose later evolution could be predominantly governed by internal secular processes.

\subsection{The Vertical Dispersion Evolution}\label{sec:sigmaz}

\begin{figure}
    \centerline{
        \includegraphics[width=1.\columnwidth]{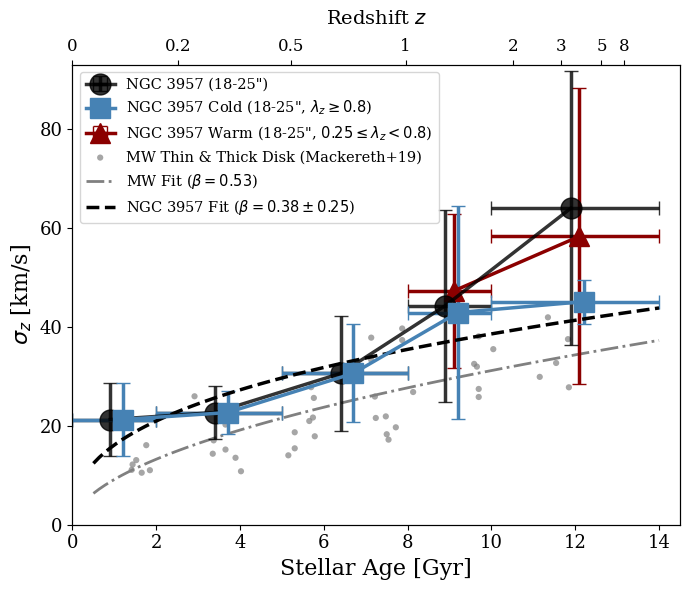}}
    \caption{
    Vertical velocity dispersion ($\sigma_z$) as a function of stellar age. The $\sigma_z$ values are calculated within a radial annulus of 18-25 arcsec (2.16-3 kpc) to minimize contamination from the central bar, bulge, and NSD. We show the kinematics for the total stellar population in this region (black circles and solid line), alongside the dynamically cold ($\lambda_z \ge 0.8$, blue squares) and warm ($0.25 \le \lambda_z < 0.8$, red triangles) components. For comparison, Milky Way thin and thick disk data (grey dots) from \citet{Mackereth2019} are included. The black dashed line and the grey dash-dotted line represent power-law fits ($\sigma_z \propto t^\beta$) to the NGC~3957 cold component and the Milky Way data, respectively. The shallower slope for NGC~3957 ($\beta = 0.38 \pm 0.25$) compared to the Milky Way ($\beta = 0.53$) indicates a significantly more quiescent assembly history.
    }
    \label{fig:sigmaz_t}
\end{figure}

To robustly quantify the dynamical heating history of the disc, we calculate the intrinsic vertical velocity dispersion $\sigma_z$ as a function of age (Fig.~\ref{fig:sigmaz_t}). Crucially, we restrict this analysis to a radial annulus of $18 < r < 25$~arcsec ($2.16 < r < 3$~kpc). This region is selected to represent the main disc while explicitly avoiding the galaxy centre (including the bar, hot spheroid, and NSD), where non-circular motions might artificially inflate the measured velocity dispersion. To accurately obtain $\sigma_z$ within this specific annulus, we project our reconstructed orbit-superposition model to a perfectly face-on view. For each stellar age bin (i.e., $0-2$, $2-5$, $5-8$, $8-10$, and $10-14$~Gyr), we directly extract the luminosity-weighted velocity dispersion of the projected pixels falling within the corresponding region. The uncertainties on $\sigma_z$ are derived by accounting for the statistical variations across all $1\sigma$ population-orbit superposition models.

In galactic dynamics, the age-velocity dispersion relation (AVR) is a fundamental diagnostic of a galaxy's assembly history and vertical heating \citep[e.g.,][]{wielen1977,Nordstrom2004,bird2013,sharma2014,martig2014,grieves2018,Mackereth2019}. 
Historically, the observed AVR was primarily attributed to pure ``vertical heating'', where stars are assumed to be born exclusively in dynamically cold, thin gas discs, followed by a gradual increase in their velocity dispersion via internal secular scattering and external perturbations. However, modern cosmological simulations demonstrate that the AVR is actually driven by a mixed mechanism \citep[e.g.,][]{martig2014}. In the early Universe, gas discs were highly turbulent and gas-rich, causing older stellar populations to be born in already dynamically hotter and thicker configurations (the ``upside-down'' formation scenario; \citealt{bournaud2009, bird2013}). Therefore, the present-day AVR could be a combination of these initially hotter birth conditions of ancient stars and their subsequent gradual vertical heating over cosmic time.
In Fig.~\ref{fig:sigmaz_t}, we show the AVR within $18 < r < 25$~arcsec for the dynamically cold orbits ($\lambda_z\ge0.8$, blue solid line), warm orbits ($0.25 \le \lambda_z < 0.8$, red solid line), and all orbits combined (black solid line). From $0$ to $10$ Gyr, the total $\sigma_z$ is entirely dominated by the cold orbits, as the warm orbits lack any young or intermediate-age stars. However, for the oldest stellar populations ($t > 10$ Gyr), the total $\sigma_z$ deviates significantly from the cold disc and shows to much higher values ($\sim60$ km/s). This divergence is driven by the presence of the warm orbits, which is exclusively composed of ancient stars with high vertical dispersion. When averaged together, this warm orbits inflate the total $\sigma_z$ at early epochs.

We parametrise the heating history using a power-law relation, $\sigma_z \propto t^\beta$, fitted exclusively to the dynamically cold orbits (blue dashed line). For stellar populations younger than $\sim 8$~Gyr (the three youngest data points), the measured $\sigma_z$ aligns remarkably well with this power-law track. 

To contextualise this heating rate, we compare our fitted $\beta$ with the Milky Way's thin and thick disc data from \citet{Mackereth2019}. We rely on the power-law slope $\beta$ rather than absolute $\sigma_z$ values because the latter is fundamentally tied to the local stellar surface mass density ($\Sigma$) and the disc scale height ($z_0$), roughly scaling as $\sigma_z \propto \sqrt{\Sigma z_0}$ \citep{vanDerKruit1981}. While the stellar surface density of the dynamically-defined cold disc of NGC~3957 in our selected annulus ($\sim 32.6 \, {\rm M}_{\odot}/\text{pc}^2$) is comparable to the Milky Way's Solar Neighbourhood \citep[$\sim 33.4 \, {\rm M}_{\odot}/\text{pc}^2$;][]{mcKee2015}, their scale heights differ ($z_0 \approx 450$~pc for NGC~3957 vs. $\approx 300$~pc for the Milky Way thin disc; \citealt{juric2008,vieira2023}). From the slope $\beta$, we find that the dynamically cold component of NGC~3957 exhibits a significantly shallower heating profile ($\beta = 0.38 \pm 0.25$) compared to the Milky Way ($\beta = 0.53$). 
In cosmological simulations, steeper AVR slopes typically correlate with more active merger histories \citep[e.g.,][]{grand2016}. However, since the Milky Way is already known to possess a highly quiescent assembly history without major mergers over the last $\sim 10$~Gyr \citep[e.g.,][]{Hammer2007, BlandHawthorn2016}, it is unlikely that the shallower slope of NGC~3957 simply implies an even more peaceful accretion history. Instead, this low $\beta$ value and the relatively low $\sigma_z$ are probably driven by a more quiescent heating history. As a lenticular galaxy, NGC~3957 is gas-poor and lacks prominent transient spiral arms \citep[e.g.,][]{Young2011, Kormendy2012}, which, along with giant molecular clouds, are known to act as the primary secular scatterers for stellar orbits in typical star-forming discs \citep{Spitzer1951, Sellwood2014}. 
Therefore, the inherent inefficiency of internal secular heating in this lenticular system, combined with a quiet external environment, might naturally explain the relatively shallow heating profile observed over the last $\sim 8$~Gyr.

However, this quiescent picture does not extend to the very early Universe. For the oldest stellar populations ($t > 8$~Gyr), the measured $\sigma_z$ deviates significantly upwards from the extrapolated cold-disc power-law fit. This upward deviation indicates that NGC~3957 might undergo a distinct period of intense vertical heating or chaotic assembly in its earliest epochs. 

This early deviation provides a crucial fossil record that links back to the galaxy's environmental history. NGC~3957 is a satellite residing in the small-to-medium NGC~4038 galaxy group \citep[LGG 263;][]{garcia1993}. In galactic dynamics, mergers and strong tidal interactions are more likely to happen in galaxy groups than in massive clusters. This is because the group's velocity dispersion is comparable to the internal velocities of the galaxies, allowing them to gravitationally bind and merge rather than simply fly by at high speeds \citep[e.g.,][]{boselli2006}. Furthermore, cosmological merger rates natively peak at higher redshifts \citep[$z \gtrsim 1$, corresponding to lookback times $>8$~Gyr;][]{fakhouri2010, rodriguezGomez2015}. Therefore, we propose that in the early, high-redshift group environment, NGC~3957 might have experienced events which heated the initial stellar populations, pushing them above the quiescent power-law track and creating the ancient warm/thick component. Following this early chaotic phase, the galaxy settled, and the subsequent disc grew and evolved in the remarkably quiescent state we observe today. This is further corroborated by the high and sustained cold disc fraction observed from $\sim 5$~Gyr ago to the present day.

\subsection{Secular disc fading and central prolonged star formation}\label{sec:age_gradient}

One of the most prominent features of the main disc in NGC~3957 is its strongly positive age gradient (younger in the inner disc, older in the outskirts; Fig.~\ref{fig:profiles}). Notably, this gradient behaves differently inside and outside the bar radius ($R_{\rm bar} = 18.5$~arcsec). Beyond the bar, the age increases steeply with radius, whereas within the bar region, the age profile flattens significantly, accompanied by an inverted (positive) metallicity gradient.

To distinguish whether this radial trend is solely driven by recent central starbursts or reflects a long-term assembly process, we recalculated the radial profiles of the cold disc by excluding stars younger than 2 Gyr (shown as blue dashed lines in Fig.~\ref{fig:profiles}). The robust persistence of the positive age slope beyond $R_{\rm bar}$ reveals that the `outside-in' suppression of star formation is a long-term evolutionary imprint, rather than a transient effect of recent star formation. 

This complex radial structure is highly consistent with the scenario of prolonged, bar-driven secular evolution. In the extended outer disc (well beyond the bar's resonances), star formation likely ceased early due to natural gas starvation. In such low-density environments, the longer dynamical shear timescales reduce the star formation efficiency, leaving the outer disc to passively fade \citep[e.g.,][]{bigiel2010}.
Concurrently, simulations demonstrate that stellar bars induce gravitational torques that efficiently redistribute angular momentum \citep{athanassoula2003}. Beyond corotation, the bar transfers angular momentum outwards, which can push gas to larger radii, lowering the gas surface density in the intermediate disc to below the star formation threshold \citep[e.g.,][]{spinoso2017, khoperskov2018}. This secular redistribution dynamically stabilises the gas against fragmentation, effectively starving the outer disc \citep{Rosas-Guevara2020}. 
In stark contrast, within the bar radius, the bar continuously funnels the remaining cold gas inwards \citep{newnham2020}. This prolonged central fuelling, combined with the strong radial mixing and orbital homogenization induced by the bar, naturally explains the flattening of the age gradient and the inverted metallicity gradient we observe at $R \le R_{\rm bar}$. Such flattening of age and chemical gradients within the bar region is a well-documented signature of bar-driven radial mixing in both observations and simulations \citep[e.g.,][]{Friedli1994, Zurita2008, Seidel2015b}.

This sustained, centrally-concentrated star formation persists even in the most recent epochs. The spatial distribution of our youngest stellar populations (0--2~Gyr) matches remarkably well with the current ionised gas morphology. As part of the GECKOS survey, \citet{elliott2026} analysed the H$\alpha$ emission in NGC~3957 and found that the ionised gas is predominantly concentrated within the central $r \lesssim 17$~arcsec (2 kpc). This spatial extent coincides almost perfectly with our 0--2~Gyr cold stellar component ($r \lesssim 20$~arcsec; Fig.~\ref{fig:Age_bins_whole} and Fig.~\ref{fig:Age_bins_disk}) and is roughly confined within the bar radius $R_{\rm bar}=18.5$~arcsec.

Interestingly, \citet{elliott2026} also reported that some H$\alpha$ gas exhibits counter-rotation relative to the main stellar disc. While such kinematic misalignment often hints at an external origin \citep[e.g.,][]{Bertola1992, Davis2011, Ristea2022}, our dynamical analysis provides further insights into the nature of this accretion event. The youngest stellar population (0--2~Gyr), which presumably formed from this gas, is dynamically very cold ($\sigma_z \sim 20$~km/s, see Fig.~\ref{fig:sigmaz_t}), and the underlying older main disc shows no signs of recent violent vertical heating.

A merger involving a substantial collisionless companion (i.e., possessing a massive dark matter halo or stellar core) would likely transfer its orbital kinetic energy to the main disc via dynamical friction, pumping significant energy into the random vertical motions of the stars. The fragility of thin discs to such collisionless heating has been extensively quantified in the literature. For instance, the analytic models by \citet{Toth1992} suggested that the accretion of a satellite with just $\sim 4-5\%$ of the total disc mass would inject enough energy to roughly double the vertical kinetic energy ($\propto\sigma_z^2$), compromising the thin disc structure. Detailed $N$-body simulations by \citet{Villalobos2008} further demonstrated that a minor merger with a 1:5 mass ratio typically heats the primary stellar disc to $\sigma_z \gtrsim 40-50$~km/s. On a more disruptive scale, \citet{Pablo2023} showed that an intermediate-mass merger (1:4 mass ratio) could perturb the orbital structure, severely erasing the dynamically-cold kinematic signature (high-$\lambda_z$ orbits) of the pre-existing disc.

The absence of such heating in NGC~3957 provides another observational constraint. The youngest stellar population maintains a remarkably low $\sigma_z \sim 20$~km/s (Fig.~\ref{fig:sigmaz_t}) alongside a highly populated cold orbital distribution ($\lambda_z \ge 0.8$, Fig.~\ref{fig:Assembly}). This dynamically cold signature perfectly corroborates the lack of recent stellar tidal debris in deep imaging \citep{jablonka2010}, collectively suggesting that the accreted material was probably highly gas-dominated. Because gas is collisional, it dissipates its orbital energy via radiative cooling rather than transferring it to the stars \citep[e.g.,][]{Hopkins2009}. This allows the smoothly accreted cold gas whether from the cosmic web, cooling from the hot halo, or a gas-rich micro-merger \citep{Algorry2014, Davis2016} to gently settle into the central potential well and form a new, thin star-forming disc, while largely retaining its decoupled angular momentum. Our findings for NGC~3957 thus provide a dynamically resolved local benchmark, illustrating how an isolated S0 galaxy might undergo a complex cycle of outer-disc fading and prolonged central star formation fuelled by gentle, dissipational external accretion.

\subsection{The hot component vs. main disc: A fossil record of galaxy assembly}
\label{sec:age_difference}

\begin{table}
    \centering
    \begin{tabular}{llc}
        \toprule
        \textbf{Component} & \textbf{Decomposition scheme} & \textbf{Age (Gyr)} \\
        \midrule
        \multicolumn{3}{l}{\textit{Definitions from \citet{Ding2023} (F3D)}} \\
        Disc & $\lambda_z \ge 0.8,r<R_{\rm e}$ & $5.9 \pm 0.6$ \\
        Hot  & $\lambda_z < 0.8,r<R_{\rm e}$ & $10.1 \pm 0.3$ \\
        \midrule
        \multicolumn{3}{l}{\textit{Definitions from \citet{jin2024} (CALIFA)}} \\
        Hot  & $|\lambda_z| \le 0.25,r<R_{\rm e}$ & $12.0 \pm 0.6$ \\
        \midrule
        \multicolumn{3}{l}{\textit{This work (GECKOS)}} \\
        Main Disc & $\lambda_z \ge 0.8, r > 7''$ & $7.7 \pm 0.5$ \\
        NSD & $\lambda_z \ge 0.8, r \le 7''$ & $6.9 \pm 0.4$ \\
        Hot & $|\lambda_z| < 0.25$ & $11.9 \pm 0.6$ \\
        \bottomrule
    \end{tabular}
    \caption{The luminosity-weighted mean ages of different dynamical components calculated using decomposition schemes from \citet{Ding2023}, \citet{jin2024}, and this work, respectively.}
    \label{tab:component_ages}
\end{table}

The age difference between the dynamically hot component and the main cold disc provides a crucial fossil record of a galaxy's assembly history. As shown in Table~\ref{tab:component_ages}, we derive the mean ages of these components using both a commonly used dynamical classification \citep[e.g., from the Fornax3D survey;][]{Ding2023} and our refined decomposition scheme.

If we apply the traditional Fornax3D criteria, defining the disc as orbits with $\lambda_z \ge 0.8$ and the hot component as $\lambda_z < 0.8$ within $r < R_{\rm e}$, we obtain a disc age of $5.9 \pm 0.6$~Gyr and a hot component age of $10.1 \pm 0.3$~Gyr. However, this broad classification intrinsically mixes the extended main disc with the younger NSD, and dilutes the hot component with intermediate-age warm orbits from the bar. 

To recover the intrinsic properties of the underlying large-scale structures, our detailed decomposition introduces a physical spatial cut to explicitly remove the NSD ($r > 7$~arcsec for the main disc) and applies a different circularity cut ($|\lambda_z| < 0.25$) for the hot component. Under this refined scheme, the absolute ages shift older: the main cold disc is $7.7 \pm 0.5$~Gyr, and the hot component is very old at $11.9 \pm 0.6$~Gyr. 

Crucially, despite the absolute age shifts caused by isolating the central substructures, the relative age difference between the hot component and the main disc remains remarkably consistent at $\Delta \text{age} \approx 4.2$~Gyr in both classification methods. This consistency demonstrates that the pronounced $\sim 4$~Gyr evolutionary gap is a robust physical feature of NGC~3957, which is insensitive to the exact phase-space definitions adopted for the decomposition.

When comparing these cleanly separated ages to those of other galaxy samples, NGC~3957 stands out as a remarkable outlier. In the CALIFA sample of field spiral galaxies \citep{jin2024}, high-mass spirals ($M_* > 10^{10.5} {\rm M}_\odot$) typically host hot components with mean ages of $\sim 6$~Gyr and discs of $\sim 4$~Gyr, maintaining a modest age difference of $\sim 1-2$~Gyr. The extreme relative age gap of NGC~3957 places it far off the typical relations seen in these field spirals (e.g., Fig.~12 in \citealt{jin2024}). We caution that differences in data quality (MUSE vs. CALIFA) and the underlying stellar population synthesis (SPS) models could introduce systematic offsets in absolute age determinations. However, such systematics are unlikely to artificially produce the massive $\sim 4.2$~Gyr relative age gap observed internally between the two components of NGC~3957.

Furthermore, compared to the cluster galaxies in Fornax \citep{Ding2023}, where ancient infallers show universally old ages for both hot components and discs ($\Delta \text{age} \approx 0$), the large age gap in NGC~3957 highlights its relatively unperturbed evolution. The disc was able to continuously form stars for billions of years after the central hot component had already been fully assembled.

Even more striking is the relationship between the mass and the age of the hot component. In the CALIFA sample \citep{jin2024}, ancient hot components ($\sim 12-13$~Gyr) are almost exclusively found in the most massive systems, where the hot component mass itself typically exceeds $10^{10} {\rm M}_\odot$. In sharp contrast, the hot component in NGC~3957 is surprisingly low-mass ($M_{\rm hot} \approx 4.41 \times 10^9 {\rm M}_\odot$, comprising only $14\%$ of the total stellar mass) despite being $\sim 11.9$~Gyr old. 

This low-mass, ancient hot core, coupled with the massive age gap ($\Delta \text{age} \approx 4.2$~Gyr), points towards a relatively quiet merger history. The old mean age of the hot component coincides with the early, chaotic assembly phase of the galaxy ($t \gtrsim 10$~Gyr, Fig.~\ref{fig:Assembly}). These clues imply that if the hot component here is a pure spheroidal bulge, it may have formed rapidly at high redshift ($z \gtrsim 2$). However, as discussed in Section~\ref{sec:result_decmp}, due to the lack of an explicit bar potential in our modelling, this $14\%$ mass fraction may also include contamination from the dynamically hot orbits of the stellar bar (i.e., the B/P structure). Consequently, we cannot fully decouple the true central bulge from the bar's vertically heated orbits, and $f_{\rm hot} \approx 0.14$ strictly serves as an upper limit for the central bulge fraction.

Nevertheless, regardless of the exact division between the true bulge and the bar, the preservation of such an extremely low bulge-to-total mass fraction strongly restricts the galaxy's merger history; galaxies with $B/T \lesssim 0.2$ are likely to have escaped major destructive merger events since at least $z \sim 2$ \citep[e.g.,][]{Weinzirl2009, Kormendy2010}. We note that recent cosmological simulations, such as TNG50, have demonstrated that galaxies can technically preserve or reform a low bulge fraction even after a major merger, provided the encounter is exceptionally gas-rich \citep[e.g.,][]{SotilloRamos2022}. However, these simulated merger events inevitably heat the system, resulting in dynamically thicker and warmer surviving stellar discs. Given the presence of the remarkably cold and extended main disc in NGC~3957 (characterised by its low vertical velocity dispersion), such late-time merger scenarios are highly unfavourable. This further reinforces the conclusion that NGC~3957 has likely escaped major destructive merger events since at least $z \sim 2$.

This scenario is further supported by the orbit circularity evolution (Sect.~\ref{sec:assembly_and_NSD} and Sect.~\ref{sec:sigmaz}), where the cold disc fraction ($f_{\rm cold}$) settles and remains broadly stable after $\sim 5$~Gyr ago, and the vertical velocity dispersion (Fig.~\ref{fig:sigmaz_t}) maintains a low profile over the last $\sim 8$~Gyr. Instead of experiencing violent structural resets, NGC~3957 likely evolved peacefully over the last $\sim 5-8$ billion years, allowing its main disc to grow and ultimately fade outside-in, driven predominantly by its own secularly evolving bar.

\subsection{Comparison to galaxies in the Fornax cluster}\label{sec:compare_fornax}

\begin{figure}
    \centerline{
        \includegraphics[width=1.1\columnwidth]{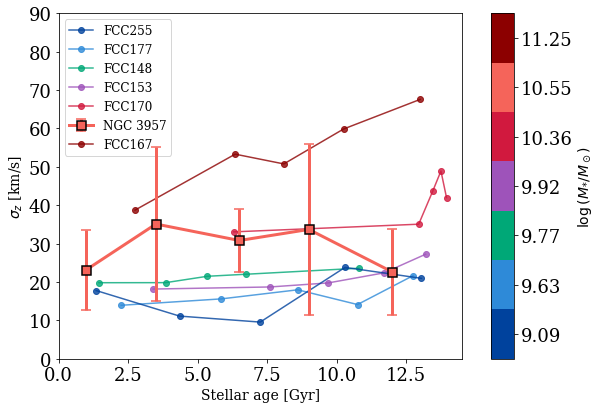}}
    \caption{Vertical velocity dispersion ($\sigma_z$) of the cold disc component as a function of stellar age. Curves are colour-coded by the stellar mass. All the selected F3D samples are S0 galaxies in the Fornax cluster. To ensure a fair comparison, the $\sigma_z$ values for both NGC~3957 and the Fornax sample are calculated identically by averaging all stars at circularity $\lambda_z>0.8$ across all available radii.
    }
    \label{fig:sigmaz_t_Fornax}
\end{figure}

\begin{figure*}
    \centerline{
        \includegraphics[width=2.\columnwidth]{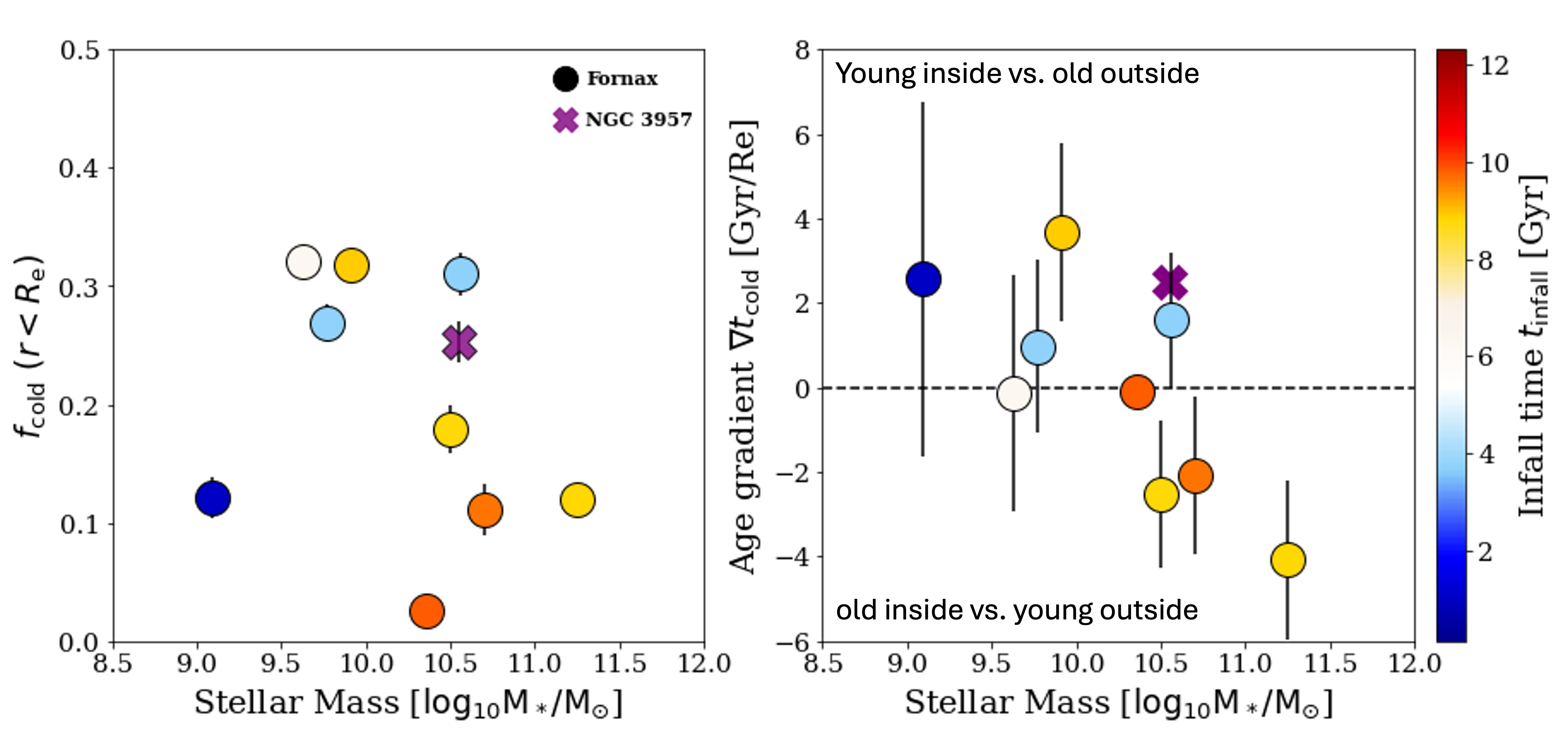}}
    \caption{
        {\em Left panel\/:}
    Cold disc fraction of the sample galaxies as a function of stellar mass. The purple point is NGC~3957. Other points are
    colour-coded by infall time into the Fornax Cluster to highlight the ancient
    (red), intermediate (white), and recent (blue) infallers. The cold disc fractions are extracted within $1R_{\rm e}$ for all the galaxies.
    {\em Right panel\/:}
    Age gradient of the discs
    as a function of stellar mass.
    The purple point is NGC~3957. Others are the Fornax cluster galaxies.
    The dashed line in the right panel marks the zero age gradient.
    }
    \label{fig:compare_fornax}
\end{figure*}

In Sect.~\ref{sec:profile}, we have demonstrated the high luminosity fraction and strong positive age gradient of the main disc in NGC~3957. In Sect.~\ref{sec:sigmaz}, we further revealed its relatively low vertical velocity dispersion within the intermediate disc region (18-25 arcsec). To place these findings in a broader context, we now quantitatively compare NGC~3957 with S0 galaxies residing in the dense Fornax cluster.

We first compare its vertical heating history with Fornax S0 galaxies using data from the F3D survey \citep{Ding2023}. Fig.~\ref{fig:sigmaz_t_Fornax} presents the $\sigma_z$ evolution for NGC~3957 alongside the Fornax sample. To ensure a strictly fair comparison, we applied the identical kinematic extraction method to all galaxies: isolating the cold disc component ($\lambda_z > 0.8$), projecting to a face-on view, and calculating the luminosity-weighted $\sigma_z$ across the entire spatial extent. 
We note that while a parallel dynamical analysis of the F3D sample exists \citep{Poci2021}, we rely exclusively on our uniformly extracted data to avoid any systematic biases introduced by different way of $\sigma_z$ extraction.

As shown in Fig.~\ref{fig:sigmaz_t_Fornax}, rather than being an extreme outlier, the vertical velocity dispersion of NGC~3957 sits comfortably within the range spanned by the Fornax S0s. Furthermore, the $\sigma_z$ values in the Fornax cluster display a clear stellar mass dependence—lower mass galaxies exhibit colder discs, while higher mass galaxies are dynamically hotter. Interestingly, NGC~3957 closely follows this mass-dependent trend, showing an evolutionary track consistent with Fornax galaxies of similar stellar mass (e.g., FCC~170). 

This finding suggests that the global vertical velocity dispersion of the cold disc is predominantly governed by the galaxy's intrinsic stellar mass, rather than the large-scale environment. Whether residing in a dense cluster or a looser group, the main disc appears to respond similarly to secular internal heating processes over time.

In the left panel of Fig.~\ref{fig:compare_fornax}, we show the luminosity fraction of the cold disc for both NGC~3957 and the Fornax cluster galaxies \citep{Ding2023}. The disc fraction and the age gradient of NGC~3957 (calculated within $R_{\rm e}$) are generally consistent with the Fornax galaxies. Galaxies in the Fornax cluster have likely experienced environmental quenching mechanisms, such as ram-pressure stripping or starvation, which progressively remove gas from the outskirts \citep[e.g.,][]{Raj2019,iodice2019,Ding2023}. 
In the Fornax cluster, these environmental effects are typically invoked to explain the `outside-in' star formation quenching and the resulting positive age gradients \citep{Ding2023,ding2024}. 

However, for a galaxy like NGC~3957 in a less dense group environment, this relatively modest central cold disc fraction and strong positive age gradient can be attributed to its specific internal assembly history. Although NGC~3957 is not in a dense cluster, its location in a small group may still eventually limit the cosmological supply of newly accreted gas. This limited gas supply, combined with a long-term internal secular evolution—such as the bar-driven inflows that continuously funnel gas into the central regions (as discussed in Sect.~\ref{sec:age_gradient}) can result in an observational signature that perfectly mimics the outcomes of environmental quenching. 

This comparison highlights a profound degeneracy in galaxy evolution: both external environmental stripping (in clusters) and internal bar-driven secular evolution (in groups/field) can result in similar `outside-in' quenching patterns, positive age gradients, and mass-dependent velocity dispersions in S0 galaxies.

\subsection{NGC~3957 as a faded spiral}\label{sec:faded_spiral}
By synthesizing the structural and kinematic fossil records extracted from our dynamical models, a coherent evolutionary picture for NGC~3957 emerges. We have independently derived multiple lines of evidence pointing toward an exceptionally unperturbed late-time evolution: 
(i) a low-mass dynamically hot core ($f_{\rm hot} \approx 14\%$) that has not been substantially grown by mergers since its early assembly ($t \gtrsim 10$~Gyr); 
(ii) a cold main disc ($\sigma_z \sim 20-30$~km/s) characterized by a shallow heating history ($\beta = 0.38$) over the last $\sim 8$~Gyr; 
(iii) a long-term positive age gradient driven by prolonged, bar-regulated secular evolution rather than a transient event; 
(iv) a dominant cold-orbit fraction ($f_{\rm cold}$) remains and settled over the last $\sim 5$~Gyr.

Taken together, these clues disfavour any major or significant minor mergers in the recent $\sim 5-8$~Gyr. While our dynamical fossil record implies a more chaotic assembly phase at high redshift ($z \gtrsim 1.5$), the subsequent billions of years of a relative quiet phase provide ample time for a star-forming spiral galaxy to gradually consume its gas and transition into a quiescent lenticular state. 

This evolutionary pathway perfectly aligns with the `faded spiral' scenario. In this framework, the morphological transformation from a spiral to an S0 is not driven by violent structural rearrangements (which would kinematically heat the disc), but by the passive exhaustion of gas (starvation) and the subsequent fading of the stellar populations in the disc. Observations of global angular momentum have confirmed that many S0 galaxies retain the dynamically cold kinematics and high spin of their progenitor spirals, distinguishing them from merger remnants \citep[e.g.,][]{vandeSande2018, rizzo2018}. 

Furthermore, as the extended spiral disc ceases star formation and passively fades over several gigayears, the young, bright stellar populations die out. As explicitly pointed out by \citet{Falcon-Barroso2015}, this passive fading of the disc naturally leads to an increase in the apparent bulge-to-disc luminosity ratio, transforming the visual morphological classification to an S0 without physically growing the bulge mass through mergers. Recent large-scale IFU surveys, such as SAMI, have further corroborated that disk fading, combined with environmental starvation in groups or clusters, is a primary driver for the formation of S0 galaxies \citep{croom2021b, croom2024}. The recent $\sim 5-8$ Gyr period of kinematic quiescence and secular settling we observe in NGC~3957 is remarkably consistent with the timescales required for this passive transformation. Therefore, NGC~3957 is very likely to serve as a dynamically resolved example of a faded spiral, whose current morphology is a direct product of long-term disc fading and internal secular evolution.

\subsection{Caveats}
\label{sec:caveats}
When interpreting the derived dark matter fraction $f_{\rm DM}$ and orbital decomposition, it is important to acknowledge the specific assumptions inherent to our dynamical modelling. A primary limitation is the assumption of a spatially constant stellar $M_{*}/L$ ratio. In standard implementations of Schwarzschild orbit-superposition methods \citep[e.g.][]{jethwa2020, Thater2022}, a constant global $M_{*}/L$ is typically adopted to convert surface brightness to stellar mass density. However, the presence of distinct radial age and metallicity gradients in NGC~3957 (Sect.\ref{sec:age_gradient}) natively implies a varying $M_{*}/L$ across the galaxy. Forcing a constant $M_{*}/L$ ratio in the presence of these stellar population gradients can introduce degeneracies between the stellar mass distribution and the dark matter halo profile, resulting in substantial uncertainties in the derived $f_{\rm DM}$.

Despite these uncertainties, several lines of evidence indicate that our global mass model remains physically robust. Crucially, during our modelling process, the dark matter halo parameters, namely the concentration $c$ and the virial mass $M_{200}$, are set completely free without enforcing any empirical mass-concentration priors. In dynamical modelling, leaving these parameters unconstrained can frequently lead to unrealistic, unphysical halo profiles (e.g., artificially cored or overly peaked halos) due to the stellar-to-dark matter degeneracy \citep[e.g.][]{Courteau2014}. However, our best-fitting values (detailed in Appendix~\ref{sec:appendix_dust}) naturally fall within physically realistic ranges expected for a galaxy of this mass, avoiding such unphysical extremes and remaining broadly consistent with the cosmological expectations for NFW haloes \citep[e.g.,][]{Dutton:2014xda}. This demonstrates the internal consistency of the model. Furthermore, the relatively low-to-moderate central $f_{\rm DM}=28\%$ (within $R_{\rm e}$) we recover corroborates the morphological features of NGC~3957. Theoretical models and cosmological simulations robustly demonstrate that the formation of strong stellar bars and subsequent B/P bulges heavily favours environments where the central regions are baryon-dominated. A highly dominant central dark matter halo would kinematically heat the disc and suppress the bar instability, preventing the buckling phase necessary to create the B/P structure \citep[e.g.,][]{Bland-Hawthorn2023, fragkoudi2025, Frosst2026}. Therefore, the presence of a prominent B/P bulge in NGC~3957 independently validates that our recovered $f_{\rm DM}$ correctly captures the true physical state of the galaxy.

A broader limitation when interpreting our derived assembly history is that our population-orbital decomposition provides a present-day snapshot of the galaxy's dynamical memory. The orbital parameters we extract reflect the \textit{current} kinematic state and locations of the stars, which do not necessarily trace their original birth environments. For instance, ancient stars that currently reside on dynamically hot orbits might have been born hot during a chaotic early assembly phase, but they could also have formed in a cold disc and been subsequently heated over time, or even been accreted from external galaxies (ex-situ origin). While combining kinematics with age and metallicity allows us to piece together a highly plausible and self-consistent formation scenario for NGC~3957, we cannot definitively rule out alternative evolutionary pathways or trace the precise kinematic origin of every stellar population.

A final, yet important, caveat concerns the treatment of the galactic bar. Our current model does not explicitly include a non-axisymmetric bar potential. However, as demonstrated by \citet{zhu2018a}, when fitting barred galaxies with such near axisymmetric Schwarzschild models, the orbits supporting the bar are broadly approximated and captured by dynamically `warm' orbits. This explains why our model can still successfully approximate the underlying potential and overall mass distribution without explicitly resolving the bar geometry. This is also the main reason why we only confidently identify three distinct dynamical components (the main disc, the hot component, and the NSD) and refrain from an overly detailed phase-space analysis of the remaining intermediate `warm' component, which may be heavily populated by these approximated bar orbits. Implementing a fully barred Schwarzschild model \citep[e.g., using the \texttt{bardisk} component in \texttt{DYNAMITE};][]{jethwa2020} \footnote{\url{https://dynamics.univie.ac.at/dynamite_docs/index.html}}, combined with advanced orbit colouring techniques, remains a promising avenue for future follow-up work to explicitly dissect the bar's orbital structure.

\section{Conclusion}\label{sec:conclusion}
We have performed a comprehensive chemo-dynamical study of the edge-on S0 galaxy NGC~3957, residing in a low-density group environment, using deep MUSE data and population-orbit superposition modelling. Our main conclusions are as follows:

\begin{enumerate}
    \item \textbf{Chemo-dynamical decomposition:} We dynamically identify three distinct components for NGC~3957. The galaxy is dominated by an extended, dynamically cold main disc $f_{\text{main disc}} = (45\pm2)\%$; $t =(7.7\pm0.5)$~Gyr; $[Z/H] = (-0.02\pm0.04)$~dex and a compact, fast-rotating NSD $f_{\text{NSD}} = (3\pm1) \%$; $t = (6.9\pm0.4) $~Gyr; $[Z/H] = (0.49\pm0.06)$~dex. We also identify a dynamically hot component $f_{\text{hot}} = (14\pm2)\%$; $t = (11.9 \pm 0.6) $~Gyr; $[Z/H] = (-0.39 \pm 0.20) $~dex that serves as the bulge.
    
    \item \textbf{NSD and bar evolution:} The NSD is the youngest and most metal-rich structural component in the galaxy. As NSDs are generally understood to form via bar-driven gas inflows, the $\sim 7$~Gyr age of the NSD suggests that the stellar bar in NGC~3957 likely formed at least $\sim 7$~Gyr ago. The NSD exhibits a strong negative age gradient, which is consistent with an `inside-out' growth scenario fuelled by secular gas inflows.
    
    \item \textbf{Main disc age gradient:} The main cold disc exhibits a strong positive age gradient (younger inside, older outside) beyond the bar radius ($R_{\rm bar} = 18.5$~arcsec). This radial trend indicates a long-term `outside-in' suppression of star formation. A plausible scenario is that the outer disc passively faded due to a gradual decline in the cosmological gas supply (starvation) or the bar-induced redistribution of gas that is pushed beyond the bar radius , while the inner regions maintained star formation for a longer period, potentially fuelled by bar-driven inflows.
    
    \item \textbf{Dynamical evolution and assembly history:} We find that the galaxy has experienced a remarkably peaceful recent history. 
    First, the circularity-age ($\lambda_z$ versus. $t$) distribution reveals that the cold-orbit fraction $f_{\rm cold}$ has remained high and stable over the last $\sim 5$~Gyr, suggesting a well-settled disc. 
    Second, the intrinsic vertical velocity dispersion ($\sigma_z$) of the main disc is exceptionally cold ($\sigma_z \sim 20-30$~km/s) across stellar populations younger than $\sim 8$~Gyr. The shallow slope of its age-velocity dispersion relation ($\beta = 0.38\pm0.25$) implies a quiescent heating history, likely precluding any significant major or minor mergers over the last 8 billion years.
\end{enumerate}

Synthesising these dynamical fossil records, specifically the low bulge mass fraction, the persistent dynamically cold disc, and the long-term positive age gradient, we find that NGC~3957 likely evolved in a relative quiescent phase following its early assembly. Its evolutionary pathway is highly consistent with the `faded spiral' scenario: as the gas supply in the group environment was gradually restricted, the extended disc passively faded, while internal secular processes (such as the stellar bar) regulated the remaining central star formation. 

Finally, a comparison with S0 galaxies in the dense Fornax cluster highlights a notable degeneracy in galaxy evolution. We find that mild gas starvation combined with bar-driven secular evolution in a lower-density environment can produce `outside-in' quenching signatures and disc fractions that are strikingly similar to those caused by rapid environmental stripping in dense clusters. To disentangle these evolutionary pathways, our upcoming work will present a comprehensive analysis of the full S0 population within the GECKOS survey, aiming to explore the diversity of their assembly histories outside of dense cluster environments.

\section*{Acknowledgements}
We thank the referee for their helpful comments.

Based on observations made with ESO Telescopes at the La Silla Paranal Observatory under programme ID 110.24AS. We wish to thank the ESO staff, and in particular the staff at Paranal Observatory, for carrying out the GECKOS observations.

This study used the Prospero high-performance computing facility at Liverpool John Moores University.

YD and MM acknowledge support from the UK Science and Technology Facilities Council through grant ST/Y002490/1.

JF-B acknowledges support from the PID2022-140869NB-I00 grant from the Spanish Ministry of Science and Innovation.

LMV acknowledges support by the German Academic Scholarship Foundation (Studienstiftung des deutschen Volkes) and the Marianne-Plehn-Program of the Elite Network of Bavaria.

FP acknowledges support from the Horizon Europe research and innovation programme under the Maria Sk\l{}odowska-Curie grant ``TraNSLate'' No 101108180.

FF is supported by a UKRI Future Leaders Fellowship (grant no. MR/X033740/1).

YJ acknowledges support by the National Science Foundation of China under Grant No. 12403017.

DAG is supported by STFC grant ST/X001075/1.

\section*{Data Availability}
The Spitzer and Pan-STARRS data ($3.6\mu $m and $r$-band images) are available to the public (Spitzer:\url{https://irsa.ipac.caltech.edu/applications/Spitzer/SHA/}, Pan-STARRS: \url{https://catalogs.mast.stsci.edu/}). The GECKOS data is available in the ESO archive.



\bibliographystyle{mnras}
\bibliography{main} 

\
\appendix
\section{MGE fitting}\label{app:MGE}
\begin{figure*}
    \centerline{
        \includegraphics[width=2.1\columnwidth]{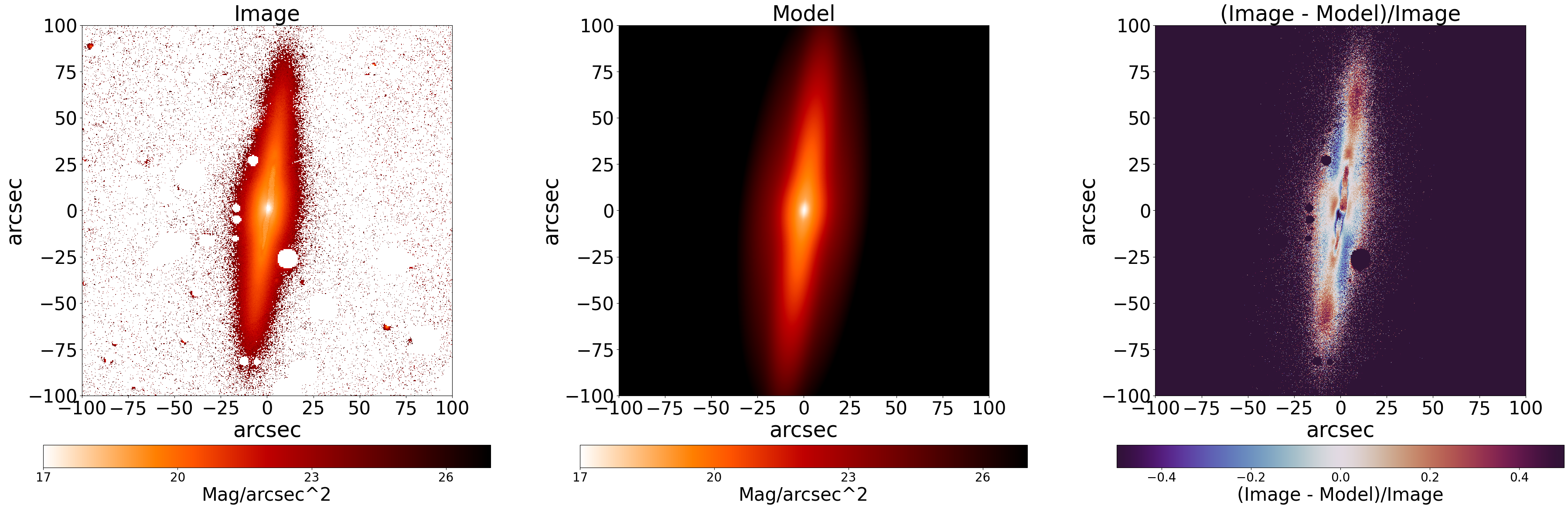}}
    \caption{MGE fit to the dust-corrected MUSE $r$-band image of NGC~3957. Left panel shows the dust-corrected MUSE $r$-band image, middle panel is the MGE fit, and the right panel is the residual between the data and fit.
    }
    \label{fig:MGE_fitting}
\end{figure*}

In this appendix, we provide the detailed parameters of the MGE models used in our dynamical analysis. Figure~\ref{fig:MGE_fitting} illustrates the MGE fit applied to the dust-corrected $r$-band image, demonstrating that the model captures the underlying surface brightness distribution well with minimal residuals. Furthermore, the specific Gaussian components, parameterised by their central surface density $\Sigma$, width $\sigma$, and axial ratio $q$, are tabulated in Table~\ref{tab:MGE_Mass} for the Spitzer $3.6\mu{\rm m}$ mass model, and in Table~\ref{tab:MGE_Light} for the dust-corrected $r$-band luminosity model.

\begin{table}
 \centering
	\begin{tabular}{c|c|c}
	 $\Sigma$ & $\sigma$ & $q$\\
	 $[\si{\Msun\per\parsec\squared}]$ & $[{\rm arcsec}]$ & \\
	\hline
    17604.47800 & 1.38840  & 0.55000 \\
    8480.60417  & 2.84854  & 0.55000 \\
    2077.29978  & 9.04445  & 0.55000 \\
    1884.79053  & 16.12154 & 0.15000 \\
    193.52096   & 29.58227 & 0.41972 \\
    1472.79852  & 31.66371 & 0.15087 \\
	\hline
	\end{tabular}
    \caption{\protect MGE parametrization of the stellar mass distribution
of NGC~3957 based on 3.6$\mu {\rm m}$ Spizter image. The central mass surface density (1), rms (2),
and axial ratio (3) of all the model Gaussians are given.}
    \label{tab:MGE_Mass}
\end{table}

\begin{table}
 \centering
	\begin{tabular}{c|c|c}
	 $\Sigma$ & $\sigma$ & $q$\\
	 $[\si{\Msun\per\parsec\squared}]$ & $[{\rm arcsec}]$ & \\
	\hline
    3718.70175 & 1.99721  & 0.54863 \\
    799.04688  & 3.57123  & 0.75000 \\
    274.07258  & 9.07336  & 0.71881 \\
    637.66074  & 17.24062 & 0.15000 \\
    290.97826  & 32.64523 & 0.15000 \\
    15.11970   & 42.19938 & 0.34175 \\
	\hline
	\end{tabular}
    \caption{\protect MGE parametrization of the stellar mass distribution
of NGC~3957 based on the dust-corrected $r$-band image. Details are same as Table~\ref{tab:MGE_Mass}.}
    \label{tab:MGE_Light}
\end{table}

\section{Robustness against dust obscuration: Mask vs. No-Mask models}\label{sec:appendix_dust}

In edge-on galaxies, equatorial dust lanes can potentially bias the extraction of stellar kinematics (e.g., by obscuring the far side of the disc or artificially suppressing the velocity dispersion). To verify that the prominent dust lane in NGC~3957 does not severely compromise our dynamical modelling, we run an additional, independent Schwarzschild model where the dust-affected kinematic bins were explicitly masked. In this section, we compare this ``masked'' model with our fiducial ``no-mask'' model presented in the main text.

\begin{figure*}
    \centerline{
        \includegraphics[width=2.\columnwidth]{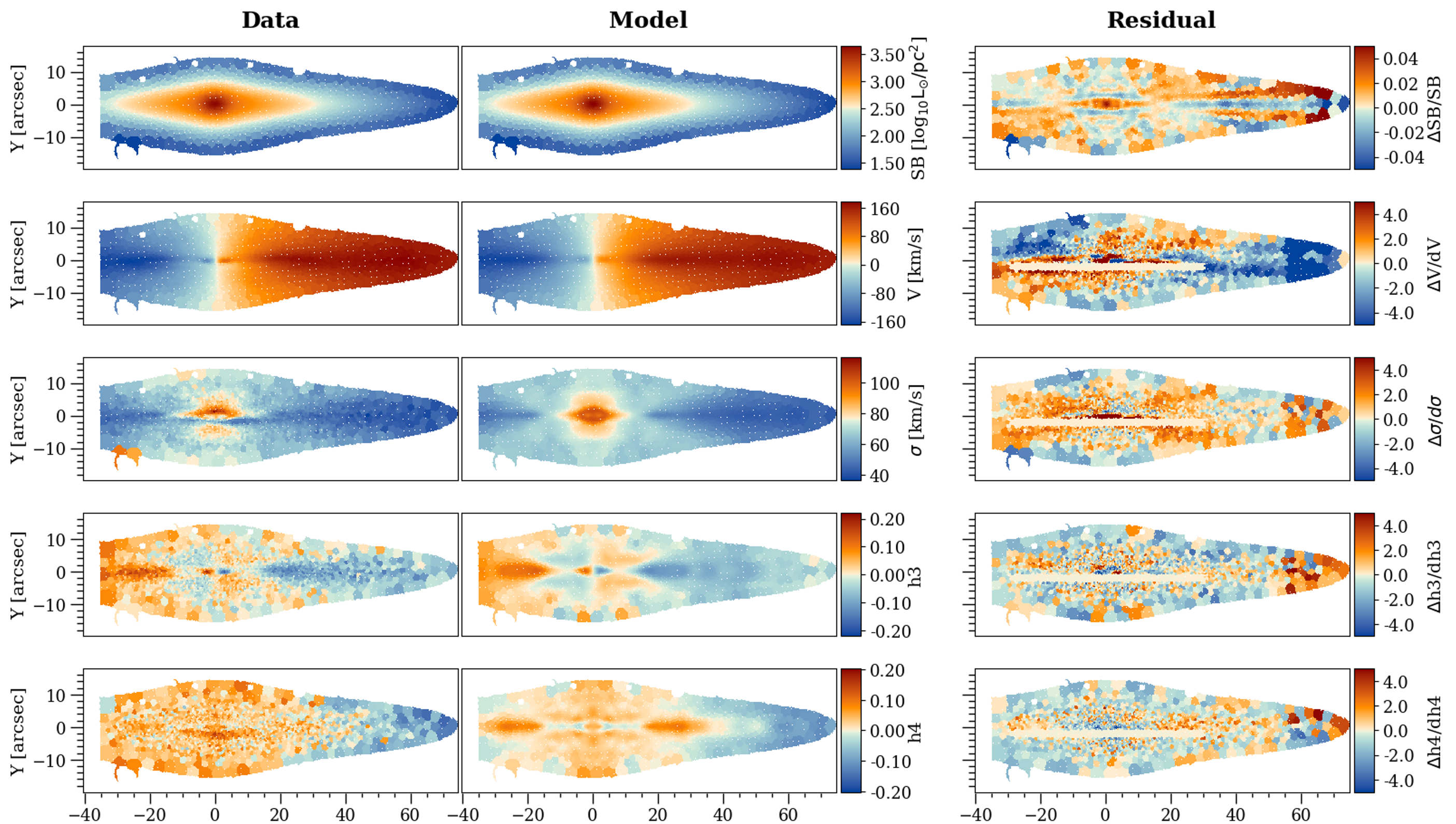}}
    \caption{
    Best-fitting population-orbit superposition model of NGC~3957 where the prominent dust lane has been explicitly masked. Similar to Fig.~\ref{fig:model_fitting}, columns from left to right show the maps of the data, model, and residuals. The masked pixels are visible as white regions in the residual maps, specifically targeting the low-$\sigma$ artifacts caused by dust while carefully preserving the central NSD kinematics.
    }
    \label{fig:model_fitting_masked}
\end{figure*}

First, we construct a tailored kinematic mask. It is important to note that this mask differs from the simple rectangular mask used for the photometric MGE fitting in Section~\ref{sec:MGE}. While we still use the colour excess map (targeting regions with $M_{3.6\mu{\rm m}} - M_{r\text{-band}} < -1.2$) as a baseline, completely masking the central equatorial region would erase crucial kinematic signatures of the NSD, particularly the fast rotation central structure in the $V$ map and the anti-correlation in the $h_3$ map. Therefore, to preserve the NSD kinematics, we apply a more localized mask. We specifically exclude the regions where the dust lane produces obvious artificial underestimation in the velocity dispersion $\sigma$ map. As a result, this tailored kinematic mask is relatively narrow in the minor axis (Y), which can be seen as the white regions in the residual maps of Fig.~\ref{fig:model_fitting_masked}.

In Fig.~\ref{fig:model_fitting_masked}, we present the kinematic maps and the best-fitting Schwarzschild model for this masked version. The model smoothly fits the remaining unmasked kinematics while preserving the overall dynamical structure. 

To quantitatively assess the impact of masking on the global model parameters, we compare the 5D hyper-parameter grid optimisation for the no-mask model (Fig.~\ref{fig:kinchi2_plot}) and the masked model (Fig.~\ref{fig:kinchi2_plot_mask}). The best-fitting parameters for both models, as well as the distributions of all models falling within the $3\sigma$ confidence level, are remarkably consistent. This indicates that the global gravitational potential and the mass-to-light ratio are not significantly biased by the localised dust obscuration.

\begin{figure*}
    \centerline{
        \includegraphics[width=2.\columnwidth]{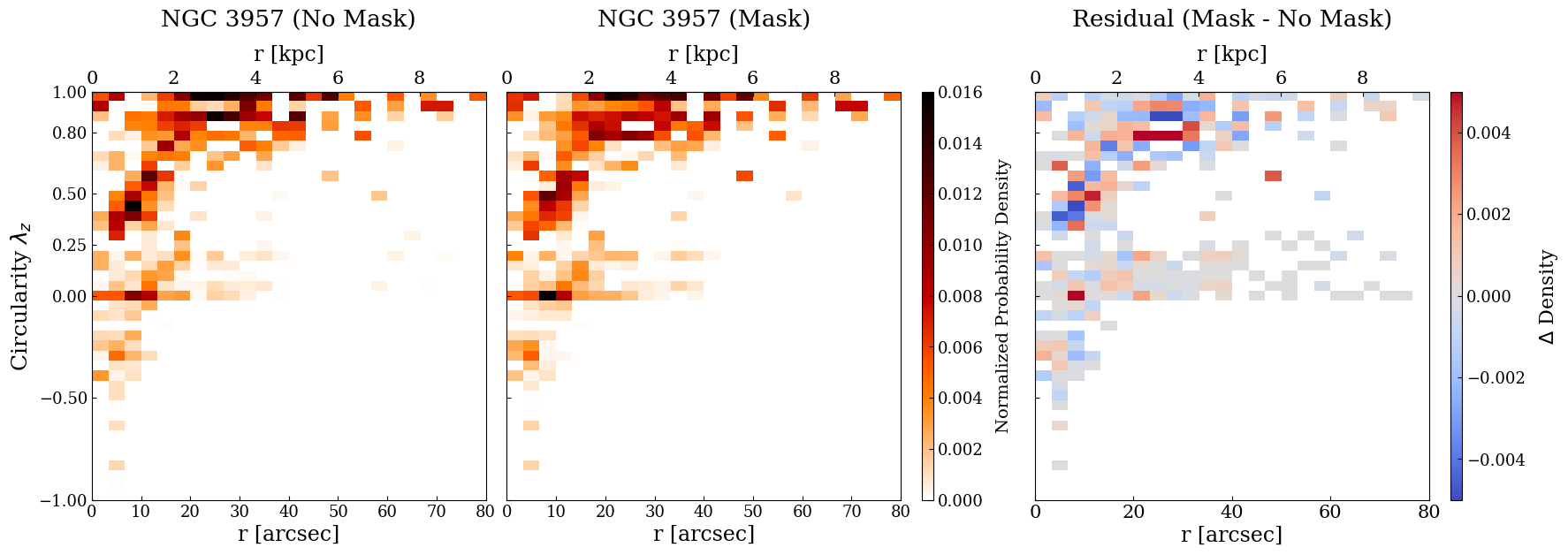}}
    \caption{
    Comparison of the phase-space orbital distributions ($\lambda_z$ vs. $r$) between the no-mask and masked models. \textit{Left:} The orbital distribution of the fiducial no-mask model used in the main text. \textit{Middle:} The orbital distribution of the masked model. \textit{Right:} The residual probability density map (Masked $-$ No-Mask). The red/blue colours in the residual map indicate regions where the masked model assigns higher/lower orbital weights. The differences are small and mostly concentrated in the dynamically warm orbits (the bar-mimicking component).
    }
    \label{fig:pps_comparison_mask}
\end{figure*}

\begin{figure*}
    \centerline{
        \includegraphics[width=2.\columnwidth]{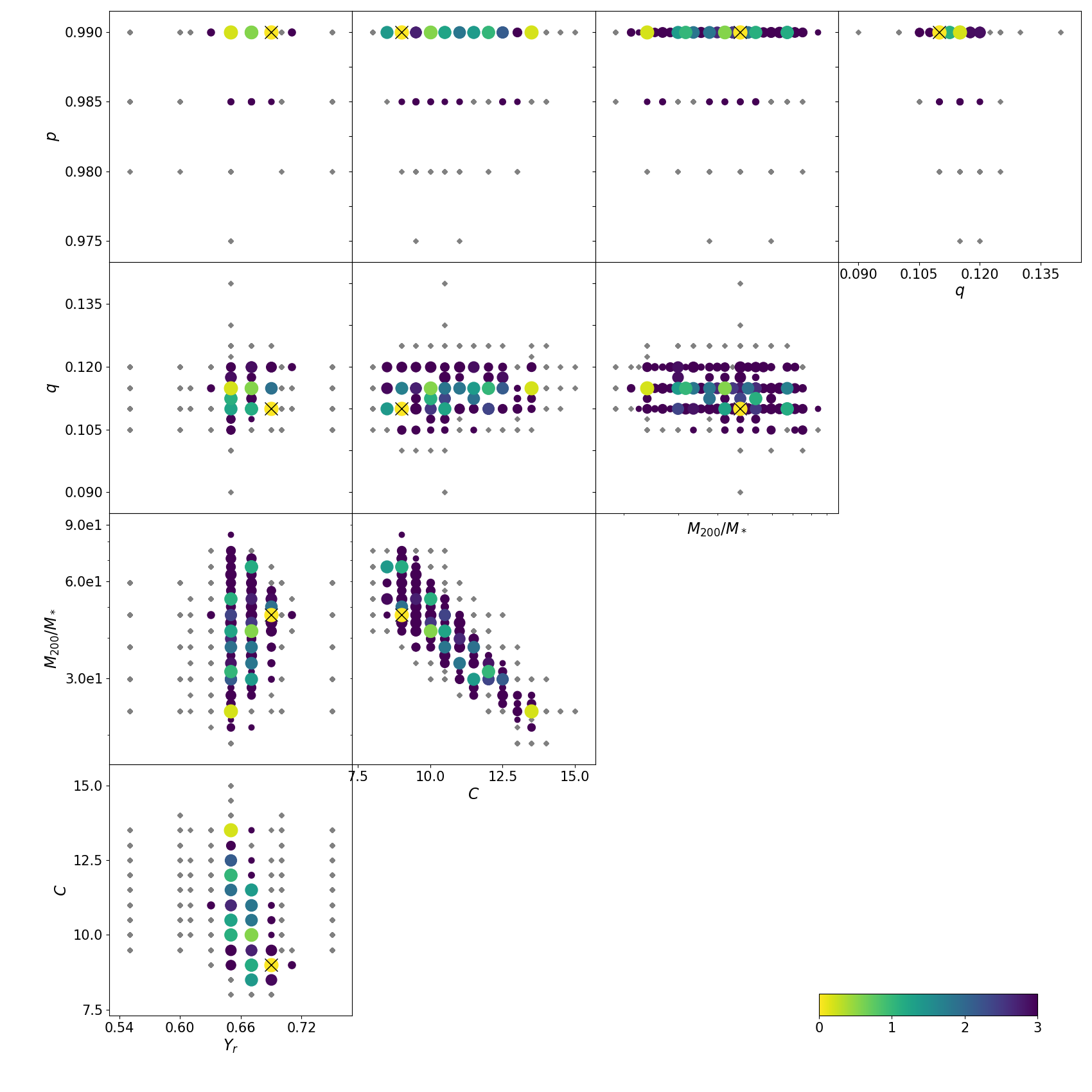}}
    \caption{
    Optimisation over the 5D parameter-space for the fiducial \textbf{no-mask} model of NGC~3957. Each point indicates an exploration of the parameter-space. The colour represents the $\chi^2$ value of each parameter set, with the best-fitting parameters indicated by the yellow-most point marked with a cross symbol. Points within the $3\sigma$ confidence level are shown.
    }
    \label{fig:kinchi2_plot}
\end{figure*}

\begin{figure*}
    \centerline{
        \includegraphics[width=2.\columnwidth]{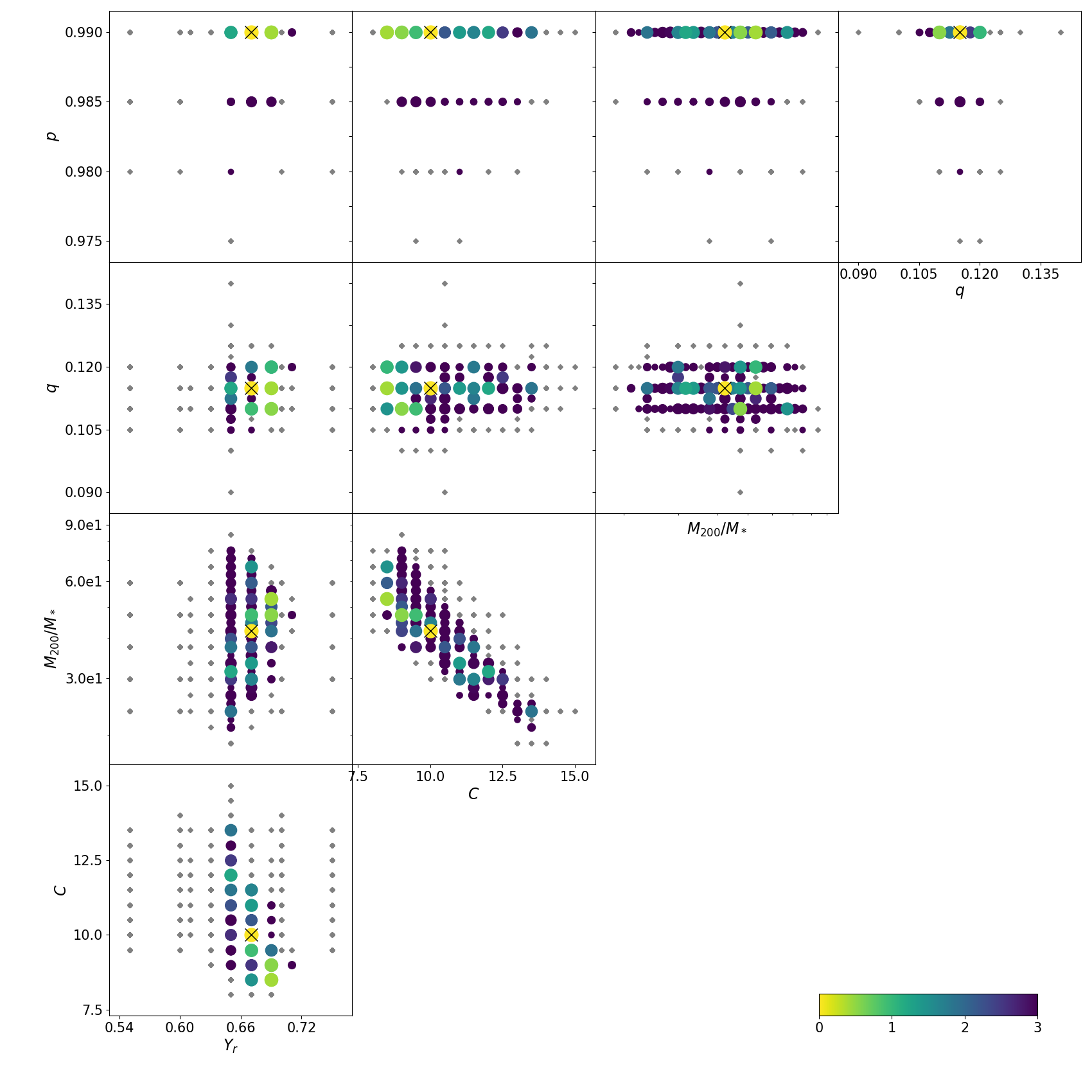}}
    \caption{
    Same as Fig.~\ref{fig:kinchi2_plot}, but for the \textbf{masked} model of NGC~3957. The distribution of the models within the $3\sigma$ confidence level and the location of the best-fitting parameter set are highly consistent with the no-mask model, demonstrating the robustness of the global dynamical fit against dust obscuration.
    }
    \label{fig:kinchi2_plot_mask}
\end{figure*}

Furthermore, we compare the intrinsic orbital distributions derived from the two models. Fig.~\ref{fig:pps_comparison_mask} displays the phase-space ($\lambda_z$ versus $r$) of the no-mask model (left), the masked model (middle), and the residual probability density between the two (right). The overall topology of the phase-space is highly consistent. From the residual map, we find that the minor discrepancies are predominantly confined to the intermediate, dynamically warm orbits ($\lambda_z \sim 0.25 - 0.8$), which correspond to the orbits mimicking the bar structure (see discussion in Appendix~\ref{sec:appendix_ramses}) . Because our primary scientific conclusions rely on the robust extraction of the dynamically cold components (the NSD and the main thin disc) and the dynamically hot component, these slight variations in the warm orbital weights have a negligible impact on our final physical interpretations.

In conclusion, the presence of the dust lane has a minimal effect on the overall dynamical modelling of NGC~3957. This robustness is physically expected, given that NGC~3957 is a globally quiescent, low star-forming S0 galaxy, where the total dust mass and obscuration are relatively modest compared to late-type starburst spirals. Therefore, adopting the no-mask model for the main analysis is reasonable and allows us to utilise the maximum number of kinematic constraints in the central region.

\section{Robustness of dynamical structure: Test with a simulated galaxy}\label{sec:appendix_ramses}

A key assumption in our dynamical modelling of NGC~3957 is the use of a gravitational potential that does not explicitly include a non-axisymmetric bar. Previous mock tests using the AURIGA cosmological simulations \citep{grand2017} showed that Schwarzschild models can recover the phase-space structure of a bar by assigning mass to dynamically ``warm'' orbits \citep{zhu2018a}. However, the galaxies in the AURIGA suite all have prominent bulges, and the simulations does not have enough resolution needed to self-consistently form a NSD.

To test our model's performance on a galaxy with both a bar and a NSD, but no bulge, we use a high-resolution, isolated hydrodynamical simulation. Testing our model against a bulgeless galaxy helps us check if our nearly axisymmetric setup might artificially create a bulge to account for complex bar motions.

\subsection{The Simulation and Mock Data}
We use a snapshot from an isolated, Milky Way-mass disc galaxy simulation run with the {\tt RAMSES} code \citep{Teyssier2002}. This simulation \citep[Fragkoudi \& Bieri, in prep, also see Appendix C of][]{Camila2023} starts with a stellar disc, a gaseous disc, and a dark matter halo, but no classical bulge. At $t \sim 3.3$~Gyr, the galaxy forms a stellar bar that drives gas inflows, building a distinct, rotation-supported NSD. We use this snapshot as mock observations by projecting it edge-on and extracting the kinematics. We then run it through our Schwarzschild modelling pipeline ({\tt DYNAMITE}) using the same setup as for NGC~3957.

To establish a ``ground truth'' for comparison, we extract the actual orbital properties of the simulation particles. We first compute the gravitational potential using the {\tt pynbody} package \citep{Pontzen2013}. Then, we integrate the particle orbits in a reference frame co-rotating with the bar ($\Omega_p = 25\, {\rm km\,s^{-1}\,kpc^{-1}}$) using the {\tt AGAMA} package \citep{Vasiliev2019}. This converts the 6D particle data into time-averaged orbital parameters (mean radius $r$ and circularity $\lambda_z$), allowing for a direct comparison with the Schwarzschild model output.

\subsection{Phase-Space and Structural Recovery}

\begin{figure*}
    \centerline{
        \includegraphics[width=2.\columnwidth]{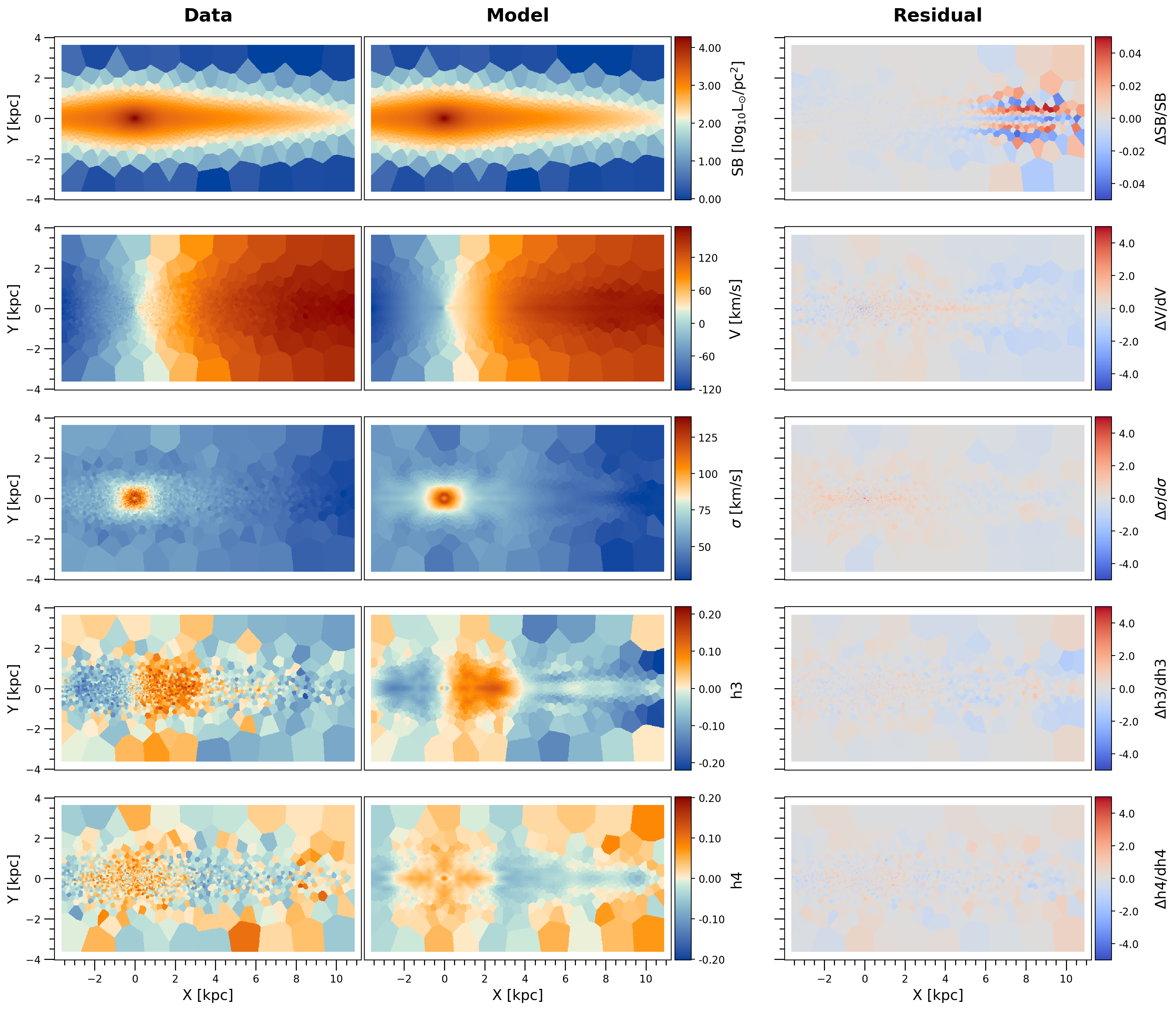}}
    \caption{
    Kinematic maps and the best-fitting Schwarzschild model without an explicit bar potential for the mock galaxy derived from the {\tt RAMSES} simulation.
    }
    \label{fig:RAMSES_best_fit}
\end{figure*}

\begin{figure*}
    \centerline{
        \includegraphics[width=2.\columnwidth]{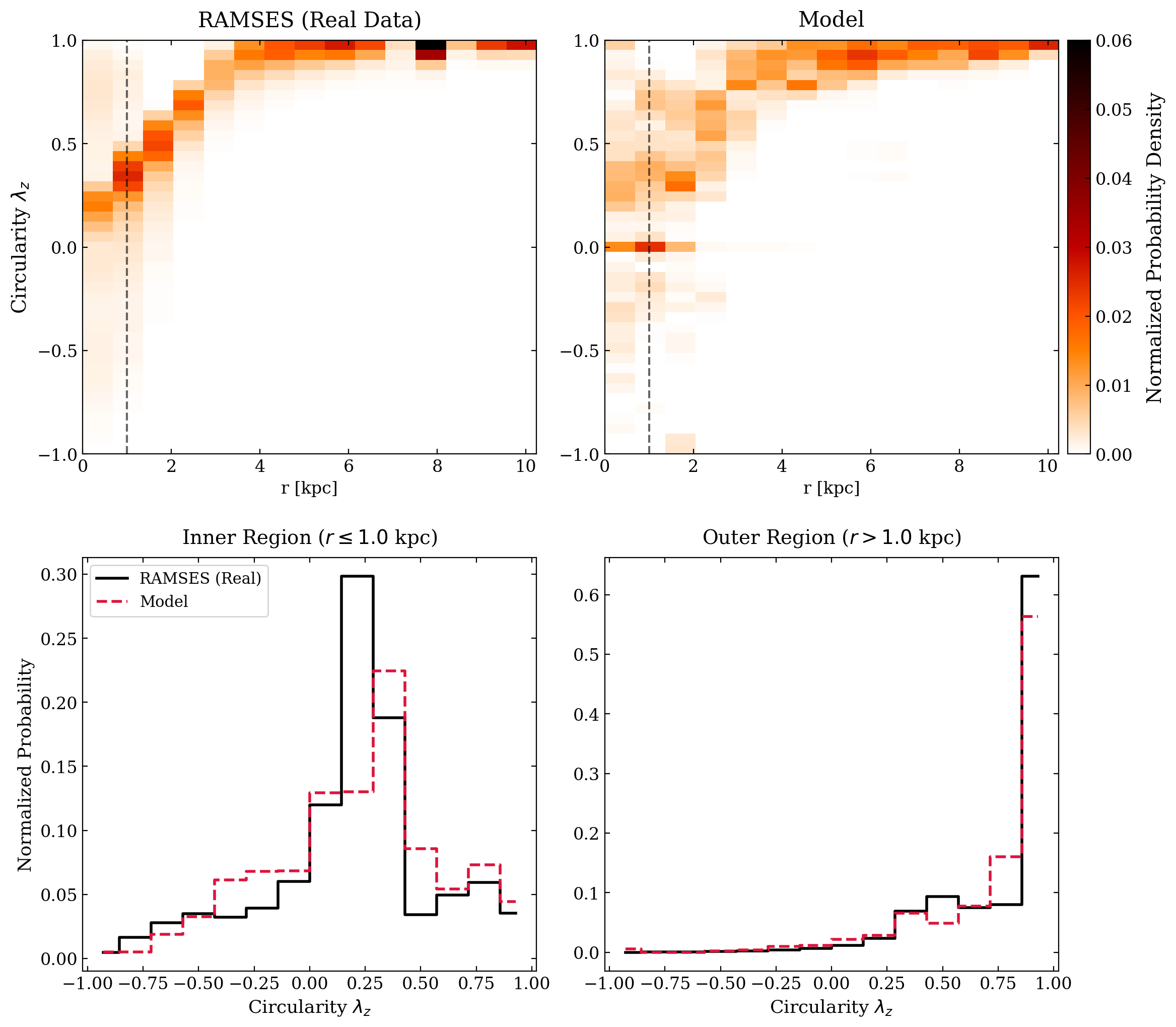}}
    \caption{
    Comparison of the phase-space distributions (circularity $\lambda_z$ vs. radius $r$). \textit{Top Left:} The true orbital distribution calculated directly from the {\tt RAMSES} simulation particles using {\tt AGAMA}. \textit{Top Right:} The orbital distribution reconstructed by our Schwarzschild model without a bar. The model generally recovers the main cold disc, the distinct NSD (high $\lambda_z$ at small radii), and the warm component representing the bar. The vertical dashed line represents the outer radius of the NSD ($R_{\rm NSD}\approx1$~kpc). \textit{Bottom panels:} The normalized 1D circularity ($\lambda_z$) distributions for the inner region ($r \le 1.0$~kpc; \textit{bottom left}) and the outer region ($r > 1.0$~kpc; \textit{bottom right}), comparing the true {\tt RAMSES} data (solid black lines) with the model reconstruction (dashed red lines).
    }
    \label{fig:pps_RAMSES}
\end{figure*}

In Fig.~\ref{fig:pps_RAMSES}, we compare the true phase-space distribution from the {\tt RAMSES} particles (left) with the reconstructed phase-space from our Schwarzschild model (right). The model captures the overall structure of the phase space reasonably well. 

We compare the mass fractions of the dynamical components using our classification criteria ($R_{\rm NSD} \approx 1$~kpc for the mock galaxy, identified from the rotation curve as in Sect.~\ref{sec:decomposition}):
\begin{itemize}
    \item \textbf{NSD} ($\lambda_z \ge 0.8, r \le 1.0$~kpc): True \textbf{1.02\%} vs. Model \textbf{1.17\%}
    \item \textbf{Main Disc} ($\lambda_z \ge 0.8, r > 1.0$~kpc): True \textbf{54.99\%} vs. Model \textbf{53.18\%}
    \item \textbf{Warm Component} ($0.25 \le \lambda_z < 0.8$): True \textbf{30.80\%} vs. Model \textbf{32.01\%}
    \item \textbf{Hot Component} ($|\lambda_z| < 0.25$): True \textbf{10.53\%} vs. Model \textbf{10.12\%}
\end{itemize}

The agreement is good. The model identifies the distinct, highly circular NSD in the inner 1 kpc (Fig.~\ref{fig:pps_RAMSES}, bottom left), suggesting that we can reasonably isolate a low-mass NSD fraction ($\sim 1\%$) from the main disc. 

We also check if the model can estimate the bar length. Using a morphological definition (where the $m=2$ Fourier amplitude $A_2$ drops by 50\%), the true bar radius is $R_{\rm bar, morph} = 3.70$~kpc. When we apply our dynamical running-window method (finding where the cold fraction $f_{\rm cold} \ge 0.5$, as in Sect.~\ref{sec:result_decmp}), the true {\tt RAMSES} phase-space gives $R_{\rm bar, true} = 3.05$~kpc, and our non-barred Schwarzschild model gives $R_{\rm bar, model} = 3.25$~kpc. These dynamical estimates are consistent with each other, indicating that the spatial extent of the warm orbits in the model can serve as a useful proxy for the bar length.

Finally, we look at the dynamically hot component ($|\lambda_z| < 0.25$). Even though the simulated galaxy is bulgeless, both the true orbital data ($10.53\%$) and the mock model ($10.12\%$) show a $\sim 10\%$ mass fraction in this region. As discussed by \citet{Tahmasebzadeh2024}, this shows a limitation of relying only on circularity to define structures: highly eccentric or boxy bar orbits often cross into the low-circularity space. Since our decomposition is based entirely on $\lambda_z$, the hot component naturally includes some of these bar orbits. The similar fractions suggest that our model recovers the mass distribution in this low-$\lambda_z$ regime reasonably well, without artificially creating a large, non-existent bulge.

Applying this insight to our results for NGC~3957, the $14\%$ mass fraction assigned to the hot component in the main text (Sect.~\ref{sec:age_difference}) likely includes some contamination from the bar's boxy or eccentric orbits. Therefore, this $14\%$ value can be viewed as an \textit{upper limit} for the actual bulge mass in NGC~3957. This is consistent with our proposed formation scenario: the bulge in NGC~3957 is likely low-mass or possibly non-existent. This points to a relatively quiet assembly history without many major mergers over the last $\sim 10$~Gyr.

\bsp	
\label{lastpage}
\end{document}